\documentclass{emulateapj}

\def\stacksymbols #1#2#3#4{\def\theguybelow{#2}
        \def\verticalposition{\lower#3pt}
        \def\spacingwithinsymbol{\baselineskip0pt\lineskip#4pt}
        \mathrel{\mathpalette\intermediary#1}}
\def\intermediary #1#2{\verticalposition\vbox{\spacingwithinsymbol
        \everycr={}\tabskip0pt
        \halign{$\mathsurround0pt#1\hfil##\hfil$\crcr#2\crcr
                \theguybelow\crcr}}}

\shorttitle{Ages of Elliptical Galaxies}
\shortauthors{Bregman et al.}

\begin{document}


\title{The Ages of Elliptical Galaxies from Infrared Spectral Energy
Distributions}

\author{Joel N. Bregman\altaffilmark{1}, Pasquale Temi\altaffilmark{2,3,4}, 
and Jesse D. Bregman\altaffilmark{2}}

\altaffiltext{1}{Astronomy Department, University of Michigan, 
Ann Arbor, MI 48109}
\altaffiltext{2}{Astrophysics Branch, NASA Ames Research Center, 
MS 245-6, Moffett Field, CA 94035}
\altaffiltext{3}{SETI Institute, 515 N. Whisman Road, Mountain View, CA 94043.}
\altaffiltext{4}{Department of Physics and Astronomy, University of Western
Ontario, London, Ontario, N6A 3K7, Canada.}

\begin{abstract}
The mean ages of early-type galaxies obtained from the analysis of 
optical spectra, give a mean age of 8 Gyr at z = 0, with
40\% being younger than 6 Gyr.  Independent age determinations are
possible by using infrared spectra (5-21 $\mu$m), which we have obtained
with the Infrared Spectrograph on the Spitzer Observatory.  This age
indicator is based on the collective mass loss rate of stars, where mass
loss from AGB stars produces a silicate emission feature at 9-12 $\mu$m. 
This feature decreases more rapidly than the shorter wavelength
continuum as a stellar population ages, providing an age indicator.  From
observations of 30 nearby early-type galaxies, 29 show a spectral energy
distribution dominated by stars and one has significant emission from the
ISM and is excluded.  The infrared age indicators for the 29 galaxies
show them all to be old, with a mean age of about 10 Gyr and a standard
deviation of only a few Gyr.  This is consistent with the ages inferred
from the values of $M/L_B$, but is inconsistent with the ages derived from
the optical line indices, which can be much younger.  All of these age
indicators are luminosity-weighted and should be correlated, even if
multiple-age components are considered.  The inconsistency indicates
that there is a significant problem with either the infrared and the $M/L_B$
ages, which agree, or with the ages inferred from the optical absorption
lines.
\end{abstract}


\section{Introduction}

Prior to the acceptance of the hierarchical paradigm for structure
formation, monolithic collapse was proposed for the formation of
ellipticals and spheroids.  In the monolithic model, an elliptical would
form at high redshift from a gas cloud that produced a brief but intense
epoch of star formation when most of the stars were formed.  A
subsequent galactic wind could keep the galaxy gas poor for most of its
life, preventing the accumulation of gas that could lead to ongoing star
formation.  This picture was in good agreement with the data \citep
{pipi03, pipi05}, yet modern theories of hierarchical structure formation indicate that
mergers should be important through cosmological time (e.g., \citealt{
kauf98}).  In the case of elliptical galaxies, this meant that existing
galaxies would merge to form larger ones, with accretion possibly
triggering a burst of star formation in the cold gas in these galaxies.  If
nearly all of the star formation were to occur early, it would imply
important changes to the basic hierarchical model, so it is essential to
have good knowledge for the age of elliptical galaxies.

Stellar ensembles that are less than 2 Gyr in age show prominent
A or F star features, enabling unique age determinations.  However, once
the age of the system is a few Gyr, the colors and metal line strengths of
the galaxy do not lead to a unique age because of a degeneracy between
the metallicity and age parameters.  That is, an old metal-poor galaxy
will have the same colors and metal line strengths as a younger metal-rich 
galaxy.  To break this age-metallicity degeneracy, investigators
measure the strength of the Balmer lines (notably H$\beta$; dominated by
turnoff stars) along with a variety of metal line indexes in the spectrum
of a galaxy (due to AGB stars), which they compare to models \citep
{worth94a, worth94b}.  There are a number of effects that can contaminate
the age-dating, but \citet{trag00a} discuss the most likely candidates and
argue that they are unlikely to cause problems.  It would seem that this
method should give a good estimate for the mean age of a single
population.

\citet{trag00a} find that for stars within r$_e$/2, the metallicities are
generally near-solar and with a small range of abundance enhancements. 
The ages, however, show a much broader range, with a median value of
7-8 Gyr and nearly 40\% of the galaxies with ages of 6 Gyr or less.  These
relatively young ages correspond to a median redshift of about 1, with
about a 40\% of the sample having formed at z $<$ 0.6.  This supports the
hierarchical model with recent merging events.  Whereas some mergers
are observed today, current surveys suggest that most of the stellar
populations in elliptical galaxies have been evolving passively since at
least z $\sim$ 1.  This is in apparent conflict with the age distribution of
Trager, so either their approach has unanticipated flaws, or multiple
populations conspire to produce the observed ages.  To help resolve this
possible conflict, an independent age estimator is required, and the
development of such a tool is the purpose of this paper.

\section{Infrared Spectral Energy Distributions as Age Indicators}

The optical age-dating technique makes use of absorption lines
that are produced in the photospheres of stars, generally at the turnoff
and on the giant branch.  In contrast, the infrared signature that we will
use is produced in the outflow of mass-losing giants, and in particular,
the asymptotic giant branch stars.  Asymptotic giant stars have very
substantial slow winds, within which the density is large enough for
material to condense, leading to grain formation.  Silicates are common
and they are formed close enough to the star to be heated to $\sim$ 300K. 
These silicates reradiate in a few broad regions, such as 9-12 $\mu$m and
another centered near 18 $\mu$m.  At shorter wavelengths (6 $\mu$m), the
continuum is produced close to the stellar photosphere, so the 5-21 $\mu$m
region contains both circumstellar emission and near-photospheric
emission.

As an ensemble of stars age, the overall luminosity of the
population decreases roughly as t$^{-1}$, so it might seem that the best
measure of age would be the mass-to-light ratio.  However, the mass is
the sum of the stellar component and the dark matter component, and one
must have confidence that the stellar component dominates this ratio in
the region it is measured.  Alternatively, one can seek to find spectral
features that have different time dependencies, so that their ratios change
monotonically with time.  For example, models indicate that the 6 $\mu$m
luminosity density decreases as t$^{-0.77}$ in solar-type populations 
(Figure 1), while the
emission from the silicate mass loss decreases more rapidly, as t$^{-1.09}$ (9.6
$\mu$m luminosity density).  The ratio L(9.6 $\mu$m)/L(6 $\mu$m) $\propto$ t$^{-0.32}$, so
there is a mapping from this ratio to the mean age of the stellar
population.

To calibrate this method, one can appeal either to models or to
stellar systems whose metallicity and age are known.  Eventually, we
hope to apply the latter method, but for this study, we used stellar
evolutionary models.  Only recently have stellar evolutionary models
included the observable consequences of mass loss from AGB stars \citep{
bres98,lanc02, mouh03} and here we use the models developed by \citet{
piov03}.  They have produced detailed spectral energy distributions from
0.1-100 $\mu$m for a single population of stars at a sequence of ages and for
three different metallicities, including the Solar value (Z = 0.02), used
here.  Their models include the effects of dust-enshrouded AGB stars,
with emission from silicates, carbon, and silicon carbide, the latter two
being important in massive stars.

We have analyzed their models, made available for use by
the authors, to determine the most suitable infrared indices that can act as
an age indicator over the wavelength region for which we have data, 5.2-
21 $\mu$m.  We find that the spectral slope, $dlnF_{\nu}/dln{\nu}$, from 5.2-6.5 $\mu$m is a
useful age indicator, as it steepens from about -0.27 at 2 Gyr to -1.07 at
16 Gyr (Figure 2).  There is scatter around this relationship due to the
finite number of stellar models used in the calculations.  That is, stellar
models are calculated for every 0.1 $M_{\sun}$ near values of 1 $M_{\sun}$, and
the turnoff ages are 9.3 Gyr for a 1 $M_{\sun}$ star, 6.5 Gyr for a 1.1 $M_{\sun}$
star, and 4.5 Gyr for a 1.2 $M_{\sun}$ star.  Not only do the relative state of
the core and envelope change in this mass range, but the dust properties
change as well, so the change in the relative fractions of these stellar
components can cause the observed non-smooth behavior in the 6 $\mu$m
slope (this is also true of the flux ratios that we will use).  Consequently,
we have fit a smooth line through these model points (Figure 2).

Aside from this slope, one can use the luminosities at 6 $\mu$m (from
the power-law fitting to the 5.2-6.5 $\mu$m continuum), the luminosity at the
local minimum of 8.3 $\mu$m (averaged 8.0-8.5 $\mu$m), and a luminosity at 9.6
$\mu$m (averaged over 9.2-10.0  that measures the silicate excess).  Of the
ratios that can be formed with these luminosities, the  L(9.6 $\mu$m)/L(6 $\mu$m)
and  L(6 $\mu$m)/L(8.3 $\mu$m) quantities are the best tracers of age.  Another
age indicator can be formed by using a luminosity at longer wavelength
L(14 $\mu$m), compared to L(6 $\mu$m).  The boundaries of these luminosities
were chosen to avoid possible PAH bands (strongest at 7.7 $\mu$m, 11.3
$\mu$m), although PAH emission is rarely seen in the galaxies we observed. 
For the following analysis, we will primarily use L(9.6 $\mu$m)/L(6 $\mu$m), as
it has the greatest range and the smallest relative uncertainty.
The mapping between this ratio and the age is 
$t = 2.6\,(L(9.6\mu m)/L(6\mu m))^{-3.1}$ Gyr.

The models of Piovan are the first at this level of detail and
cannot be expected to be perfect, so the absolute calibration of the
models may be somewhat incorrect.  However, the relative trend should
be reliable in identifying old and young systems.

\section{Sample Selection and Observational Program}


Our sample is taken from a survey of 51 nearby ellipticals for which
high-quality optical spectra were obtained (Trager et al. 2000, and
references therein). Although ages derived from absorption line indices have
uncertainties associated with the adopted stellar models and  patterns
of abundances when alpha-enhanced mixtures are adopted (Tantalo
\& Chiosi 2004), these galaxies have well established age-metallicity
determinations among nearby elliptical galaxies. Others have determined
ages for nearby ellipticals (summary by Terlevich \& Forbes 2002; recent
determinations by Thomas et al. 2005a), which are similar and are discussed
further below.

From the \citet{trag00a} sample, we have
taken a subset of galaxies with the properties:  
B$_{0T}$ $<$ 12.5; M$_B$ $<$ -19 (no
dwarfs); and no AGN activity (two 3C objects are excluded). This leads
to a well-defined sample of 31 galaxies covering a broad range of
implied ages, from 2 Gyr to 15 Gyr, with a fairly even distribution by
age:  8 galaxies have estimated ages of 2-5 Gyr, 7 are between 5-8 Gyr, 9
are between 8-12 Gyr, and 7 are between 12-18 Gyr (Table 1; the metallicity
Z$_H$ and ages are within r$_e$/8 as given in \citealt{trag00b}).  The mean
iron metallicity for the sample is [Fe/H] = 0.06 and the 25/75\% quartiles are -
0.04 to 0.15, so the range is modest. This sample is representative of
elliptical galaxies with optical luminosities near L$_*$.  For one galaxy, NGC 3608, there were
significant instrumental difficulties and the resulting data are not reliable,
so it is not included.  For one other galaxy, NGC 4697, the spectrum is
completely different than the rest, is not useful for age determinations,
and will be discussed elsewhere.

Based on the spectral model of Piovan (and observational data of
AGB stars), it is important to define both the underlying continuum as
well as the silicate feature, and this requires spectroscopic coverage from
about 5-20 $\mu$m.  
This type of coverage was realized with the combination of
Short--Low, and Long--Low modules on the Infrared Spectrograph (IRS;
\citealt{houck04}) on board the {\it Spitzer Space Telescope}
\citep{wern04}. These modules together provide spectral coverage of the 
region 5.2-21.3 $\mu$m with a spectral resolution 
$R= \lambda / \Delta \lambda$ of 64-128, 
depending on the wavelength.  Both segments of the Short--Low 
spectral region (5.2--7.7 $\mu$m for the SL2 module, 7.4--14.5 $\mu$m 
for the SL1 module) use
the same slit size (3.7\arcsec\ by 57\arcsec) and position angle, but the
Long--Low (14.0--21.3 $\mu$m for the LL2 module) slit is 10.5\arcsec\ by 168\arcsec\ and is at a
position angle about 90\degr\ different than the Short--Low.  In this case,
we will match the spectra by using the overlap region at 14 $\mu$m.

The galaxies are not point sources, but are well-contained within
the slits as the typical half-light radius is 30\arcsec.  In the IRS staring
mode, two spectra were obtained per setting, with each offset from the
center of the slit by one third of a slit length.  This results in the endpoint
of the Short-Low slit in the off position is 38\arcsec\ from the center of 
the galaxies where the
galaxy signal is about 2\% of the signal at the center, or 0.2 mJy on
average (much smaller than the measured signals).  For the Long-Low,
the offset is much greater, so the contribution from the galaxy in the off
position is negligible.  The integration time for each galaxy is the same,
consisting of 8 cycles of 14 sec ramp time for Short--Low and 6 cycles at 30
sec for Long--Low. 

\section{Data Processing}

The standard IRS data reduction pipeline, version S12.0, at the Spitzer 
Science Center (SSC) was used to reduce the data. This reduction includes ramp
fitting, dark-sky subtraction, droop correction, linearity correction, 
flat-fielding, and wavelength and flux calibration \citep{decin04}. 
Before performing the spectral extraction, the local background for the SL
modules was subtracted using observations when the target was located 
in an alternate slit. At longer wavelengths, since we recorded data using only 
the LL2 module, the local background was subtracted by differencing the 
two nod positions along the slit.
The spectra were then extracted from the sky-subtracted two-dimensional 
array images using the SMART (ver. 5.5.6) software package \citep{higd04},
after the mean of the flux estimates from each ramp cycle were
combined.
We performed a "fixed column extraction" because the emission from each
galaxy did not resemble the point source profile, showing substantial 
extended emission. 
In its current form, the SMART software is optimized to perform 
spectral extraction and flux calibration for point source targets.
In order to correct
for the use of the standard flux conversion tables, which are based
on point source extraction, we applied a
correction that accounts for the aperture loss due to the
narrowing of the extracting column as a function of wavelength used by 
SSC and that in turn feeds back into the FLUXCON tables.
Observations of the standard star HR 6348 were used to calibrate our
target spectra. A spectrum of HR 6348 was constructed combining a number
of observations recorded in 2004 under the program ID 1404.
Our spectra were calibrated by dividing the extracted spectrum of the
source by the spectrum of the standard star, extracted with the same
extraction parameters applied to our target sources, and multiplying
by its template \citep{cohen03}.

For sources with extended emission and spatial structure, the 
spectrum is the convolution of the 
source brightness distribution with the beam profile.
In such a case a rigorous flux calibration requires 
a reliable model of the source geometry, and
a characterization of the instrument's beam profile.
Since the beam profile, defined as the map obtained by moving
a point-like source across the aperture and its variation with
wavelength is not publicly documented yet, we did not apply
any correction due to diffraction losses or gains in the slit
that are inherent to extended sources.
Because the wavelength dependence of such a correction may cause
an additional slope in the observed spectra \citep{smit04}, we 
compared the extracted IRS fluxes with the measured IRAC and MIPS
density fluxes for those galaxies that are listed in our sample and 
have broad-band observations available in the {\it Spitzer} public archive.
Figure 3 shows the low 
resolution spectra of seven elliptical galaxies with data points 
from the four channels of the IRAC instrument and the MIPS 24$\mu$m channel. 
For each galaxy, the broad--band photometry, performed
within a aperture equal to r$_e$/8, centered on the galactic nucleus, has 
been scaled to match the IRS spectrum at 8 $\mu$m.
The spectral slope in the 4--8 $\mu$m region, as derived by IRS and IRAC 
measurements, is in good agreement,
giving us confidence in the accuracy of the extracted spectra; any correction 
due to diffraction losses is likely to fall inside the calibration 
errors of IRAC and MIPS observations that are of the order of 20\%.

Fluxes and slopes that have been extracted from these spectra (shifted
to z = 0) are given in Table 2, where the slope at 6 $\mu$m is given in 
column 3, and the wavelengths ($\mu$m), fluxes (mJy) and uncertainties 
(mJy) are given at four wavelengths.  The uncertainties in the fluxes are 
only due to the statistical errors obtained in the data reduction process.
The individual spectra, without the small redshift corrections, are shown
in Figure 4.

\section{Results and Interpretation}

The results of the infrared age indicators are discussed first,
followed by a comparison with the ages that one infers from the observed
values of $M/L_B$.

\subsection{Ages from the Infrared Data}

We begin by inspecting a single typical spectrum, NGC 584, which is
compared to a single-epoch stellar population model (Figure 5).
The models with ages of 5 Gyr or less are clearly ruled out and
the 5.3-9.6$\mu$m continuum is best fit with the 14 Gyr model.
None of the models are particularly accurate in reproducing the 
10-21$\mu$m continuum, but the 14 Gyr spectrum appears to be the best.
Narrow atomic emission lines from [Ne II lambda 12.8$\mu$m] and [Ne III
lambda 15.3 $\mu$m] are common features and will be discussed separately

The other important spectral features that occur in this
wavelength range are from PAH emission, where the strongest lines are
usually at 7.7 $\mu$m and 11.3 $\mu$m.  One galaxy, NGC 4697, shows PAH
emission that dominates the spectrum, making it nearly impossible to
remove from the underlying stellar spectrum; this galaxy is excluded
from study here but will be discussed elsewhere.  The only other galaxy
with detectable PAH emission is NGC 3379, but the lines are weak so
the stellar spectrum can be analyzed.  The 7.7$\mu$m feature can contaminate
the L(6 $\mu$m)/L(8.3 $\mu$m) ratio, which may have occurred in the case of
NGC 3379.

For the 29 galaxies in the sample, their spectra show tremendous
similarity (Figure 4).  There is
no obvious difference in the spectral energy distribution between
galaxies that have young or old ages according to \citet{trag00b}.  For
example, none of the galaxies in the sample are similar to a 3 Gyr
population, all appearing to be older populations (Figure 5).
To quantify this, we use the L(9.6 $\mu$m)/L(6 $\mu$m) ratio as an age 
indicator, as
it has the greatest leverage on the age, is insensitive to the precise bands
used to define the ratios, and has a typical uncertainty of 5-10\%. 
The distribution of this ratio has a mean value of 0.580 $\pm$ 0.008, where
sigma = 0.041 and for quartile points of 0.55-0.66 (Figure 6).  According to the
models of Piovan, this would imply a mean age of 13 Gyr and a range
defined by the quartiles of 9-16 Gyr (Table 3).  
We are not confident that the models of Piovan have sufficient
absolute accuracy, as some of the deduced ages are greater than the
age of the universe, 13.5Gyr for 
$\Lambda$-CDM universe with H$_0$ = 70 km$^{-1}$ s$^{-1}$ Mpc$^{-1}$.
This is due mostly to the many uncertainties
still affecting the modeling of the low-mass AGB stars, which, at
older ages, dominate the AGB population.
\citet{temi05b} found a similar offset in the absolute calibration 
of the Piovan models when comparing broad--band mid-infrared 
ratios with the predicted flux ratios for a small sample of elliptical 
galaxies.
If we were to shift the
scale of Piovan such that the mean age is 10 Gyr, the range in L(9.6
$\mu$m)/L(6 $\mu$m) implies an age range of 8-13 Gyr.

The inferred age range seems moderately restricted, but it is likely
to be narrower since the uncertainties introduced by measurement error is
similar to the sigma inferred from the distribution (see below).  When the
L(9.6 $\mu$m)/L(6 $\mu$m) age indicator is compared to the ages of Trager, there
is no correlation (Figure 7).  This is to be expected if the
values of L(9.6 $\mu$m)/L(6 $\mu$m) are random scatter about a mean value, but
it does pose a conflict between these two age indicators, discussed
below.  Finally, we find that there is no correlation between L(9.6
$\mu$m)/L(6 $\mu$m) and $M/L_B$, which also is explained
if the galaxies all have similar ages.

Potentially the second best of our infrared age indicators is the 6 $\mu$m spectral
slope (Figure 8).  The mean value is -0.95 and nearly all of the
galaxies are within 2$\sigma$ of this value, so a considerable amount of the range
is due to measurement error.  The conversion from slope to age may be somewhat
incorrect since a slope of -0.95 corresponds to 14 Gyr, which is slightly greater
than the age of the $\Lambda$-CDM universe.  Assuming that the slopes are
dominated by photon statistics, the observed distribution requires variation
in the age distribution.  If the true mean age is assumed to be 10 Gyr and the
underlying age distribution is a Gaussian, its corresponding $\sigma \simeq$ 3 Gyr.
As with the above infrared age indicator, there is no correlation with the
ages of \citet{trag00a}.

\subsection{M/L as an age indicator}

The property of a single stellar population that changes the most
with age is the luminosity, so we can examine whether the ratio M/L
contains useful age information and if the optical or infrared age
indicators are related to M/L.  The choice of L is not particularly
important since most decline with age as t$^{-1}$, but here we choose L$_B$,
where the models show that L$_B$ $\propto$ t$^{-0.95}$ in the age range of
interest \citep{bruz03}.  The mass within r$_e$ is proportional to r$_e$ $\sigma^2$, and even the
very thorough determination of the mass by the SAURON effort \citep{capp05} finds that
this is a reasonable quantity to use (the difference in the mass between
the two methods is small compared to the range of M/L in the sample). 
We use the value of $M/L_B$ as given by \citet{trag00b} but since this
quantity is known to increase as L$_{B}$$^{1/4}$, we correct this quantity for
this effect, normalizing the values of $M/L_B$ at B$_{0T}$ = -21 (this reduces 
the scatter in the ensuing figures).  
A plot of $M/L_B$ vs the Trager ages shows a
modest correlation, but a large amount of scatter (Figure 9).  The data are
inconsistent with the time evolution of a single-age population.

Another approach for obtaining ages is to assume that the shape
of the mass distributions is similar between galaxies so that $M/L_B$
translates into an age.  For this approach to yield correct results, the dark
matter to luminous matter ratio should be either small or constant, which
appears to be the case \citep{kron00, sagl00, thom05b}.  Also, a correct age
determination would require that $M/L_B$ should be dominated by a
single age population.  Inferring ages from their $M/L_B$ values (using the
calibration of \citealt{bruz03}), the
galaxies cluster in age at 8.0-12.4 Gyr (25\% to 75\% quartile ages) with a
median at 9.4 Gyr and an effective sigma for the distribution of 2.8 Gyr
(Figure 10; Table 3).  The quantity $M/L_B$ depends on r$_e$, $\sigma^2$, M$_B$,
metallicity and distance d, which each have an uncertainties and when
added incoherently, leads to a typical uncertainty in an individual $M/L_B$
of 25\%, which would introduce an uncertainty in the age of about 2.5
Gyr.  Therefore, much of the range of inferred ages may
be related to observational uncertainties in determining $M/L_B$.  
There are a few galaxies with ages near 20 Gyr, one of which is the well-
studied system NGC 1399, that is was also observed by \citet{sagl00}. Using
a detailed dynamical model, they found $M/L_B$ = 10, which corresponds
to an age of 13 Gyr.  This points out that these large ages can be due
to the uncertainties of the method in determining $M/L_B$.

\subsection{Are there any young galaxies in this sample?}  

We have searched for the most likely young galaxies by
examining those that have the youngest ages as given by \citet{trag00b}
and \citet{thom05a}, while also having low $M/L_B$ and 
high L(9.6 $\mu$m)/L(6 $\mu$m) ratios (Table 2, 3).  Two
galaxies fit that category, NGC 3377 and NGC 1700.  For NGC 3377,
Trager list an age of 3.7 Gyr, $M/L_B$ = 4.3 (5.3 $\pm$ 1.3 Gyr), and L(9.6
$\mu$m)/L(6 $\mu$m) = 0.67 $\pm$ 0.07 (9 $\pm$ 2.5 Gyr).  This galaxy is a relatively
rapid rotator, so the mass may have been underestimated, which would raise
$M/L_B$ and the inferred age.  There are no obvious signs of a recent
merger in the structural properties of the galaxy \citep{schw92}, and while the infrared
ratio is higher than average, it is within 2$\sigma $ of the mean and its
nominal value, 9 Gyr, does not imply a young age.  The evidence is not
compelling for NGC 3377 being a young galaxy.

The other system, NGC 1700, has one of the youngest ages listed
by Trager, 2.3 Gyr, and values for $M/L_B$ and L(9.6 $\mu$m)/L(6 $\mu$m) of 4.4
and 0.66 $\pm$ 0.07; also, the other two infrared ratios indicate a galaxy
younger than the sample average.  The age inferred from $M/L_B$ is 5.5
$\pm$ 1.4 Gyr and the age inferred from the infrared ratio is 9.3 (-1.5/
+ 3; it would be 1.5 Gyr younger if we were to change the normalization so
that the median galaxy age is 10 Gyr).  The 3$\sigma $ lower limit on
the age from the infrared indicator is 4.8 Gyr (4.0 Gyr for the 10 Gyr
median galaxy normalization).  Other observers also suggest a young
age, but not as young as the Trager value.  \citet{stat96} suggests an age of 3-6
Gyr based on the extent of the region of relaxed dynamics near the center,
while \citet{schw92} give an age of 6 Gyr based on the galaxy color.  
The study of the globular cluster population \citep{whit97} 
shows that it is typical of normal old ellipticals, where the population lies
between the predictions for an old, metal-poor population (15 Gyr) and a
solar metallicity population (5 Gyr).  Using the same data from {\it HST} and
ground-based data, \citet{brow00} show that the ages could be as low as
3 Gyr if the globular clusters have super-solar abundances, [Fe/H] = +0.5.
The galaxy has a counter-rotating core and there is evidence of dust 
in the central region, also consistent with a merger, although the magnitude
of the merger is poorly constrained.
On balance, it seems likely that this is an
intermediate age galaxy (4-8 Gyr), but it is unlikely to be as young as
Trager suggests and it could be old, like the other galaxies.

\section{Discussion and Conclusions}

The ages deduced from the infrared spectra indicate an old mean
age (10-13 Gyr) with a narrow age distribution of only a few Gyr.  This is
consistent with the ages inferred from the $M/L_B$ distribution, as well as
for the broadband flux age indicators \citep{temi05a, temi05b}.  These results
would appear to be in conflict with the ages deduced by Trager from
their optical line indices.

The ages determined here and by Trager are single-population
ages, so we examine whether both methods are affected differently by
multiple-age populations, in which case it would be possible to identify
the contributions from multiple populations.  There are an infinite
number of ways of combining populations, so for simplicity, we use two
populations, one being old (12 Gyr) and one being younger (1 Gyr).  For
this combination of populations, we calculate the effective age if one
interpreted the result as a single-age population.  For the optical line
index ages and $M/L_B$, we use the on-line model of Worthey with Solar
metallicities while for the infrared ages, we use the Solar metallicity
Piovan models.  As the fraction of the younger population is increased,
the inferred age naturally decreases, with the $M/L_B$ indicator least
affected by the addition of a younger component, and the H$\beta$ age
indicator being most strongly affected (Figure 11).  The apparent
age decreases by a factor of two with the addition of the 1 Gyr stellar
component of 2.2\%, 4.3\%, and  9.8\% (total mass) for the age indicators
of H$\beta$, infrared, and $M/L_B$, respectively.  
In order for the inferred age
to be greater than 10 Gyr, the young population can only account for
0.52\%, 0.87\%, and 2.1\%  of the total mass of the galaxy for the  H$\beta$,
infrared, and $M/L_B$ age indicators.  
This is also consistent with the required mass fraction of young stars
($\sim 1\%$) as derived by the observed flux ratio F(24$\mu$m)/F(3.6$\mu$m)
when theoretical mid-infrared SSP models are combined \citep{temi05b}.
A remarkable implication is that
when the H$\beta$ indicators show a old age (e.g., 13 Gyr), it suggests that
there has been virtually no star formation during the recent lifetime of the
galaxy.

The general similarity between these three luminosity-weighted
age indicators implies that there will be a strong correlation between the
ages determined from the three methods.  As discussed above, the
uncertainties in the ages derived from the infrared and $M/L_B$ methods
are responsible for most of the observed range, so the lack of a
correlation between the two methods is not surprising.  However, the
range of ages from the H$\beta$ age indicator is significantly larger than the
errors, so finding a statistical correlation with the other quantities should
be possible.   There is no correlation of the H$\beta$ age indicators with the
infrared indicators and the correlation with $M/L_B$ is quite weak and of
the wrong slope.  
Maraston \& Thomas (2000) have shown that the  addition of a small old
metal-poor population can reduce the Hbeta ages of galaxies, but it is unclear
if the addition of this type of population can reduce the ages to younger values
given by Trager.
We conclude that the introduction of a second, younger
population will not lead to age determinations that are consistent between
the H$\beta$ method and either the infrared method or the $M/L_B$ method.
The discrepancy with the Trager ages remains, indicating that either the
infrared and $M/L_B$ methods are flawed, or their approach is flawed. 
Trager have considered the prime candidates for contamination of H$\beta$
and shown them to be unimportant, or possible to correct for.  Therefore,
if their method is incorrect, there would have to be a very significant
contamination channel that is not yet recognized.  The problem does not
lie with the method, because independent studies using line indices yield
similar results \citep{deni05a, deni05b}.

Insight into the formation ages of ellipticals can be obtained from
studies of galaxies at higher redshifts, although the results are not
entirely clear.  There is broad agreement that at redshifts beyond z=1, and
even z $>$ 2, massive red (early-type) galaxies have little star formation
present, indicating very old populations \citep{dadd05a, dadd05b, labb05, capp05}. 
For reference, redshifts of 0.5, 1.0, 2.0, and 3.0 correspond to ages of 5.0,
8.6, 10.3, and 11.4 Gyr in a $\Lambda$-CDM universe, while the age of the
universe is 13.5 Gyr.  Less massive galaxies appear to have more star
formation \citep{treu05}, and if this aspect remains in the galaxies today, it
might be seen in present day data.  However, we find no correspondence
between the Trager ages (or our IR ages) and the mass of our sample
galaxies (Figure 12).  Merging is likely to have occurred since z = 1 because the
number of early-type galaxies has doubled since that time.  In order not
to produce galaxies that are too blue, star formation probably was not a
significant part of these mergers \citep{bell04,bell05,fab05}.  It is still
unclear whether this degree of merging would destroy rather tight
correlations, such as between velocity dispersion and Mg line strength. 
Finally, there needs to be an analysis as to whether the age distribution
found by Trager is consistent with the observations at higher redshift and
the expected mergers.

The calibration of the infrared spectral energy distributions is one
of the important aspects of our technique that needs to be improved. 
Presently, we rely on the models of Piovan, which despite their
sophistication, are the first to predict the level of detail that we utilize. 
As with all models, there are significant improvements that can and will
be made in the future, and at some point, we hope that the data and
models are calibrated well-enough that we can make credible maximum
likelihood fits between the two.  As a complement to using improved
models, we hope to use the infrared spectra of globular clusters with
known ages and metallicities as the basis set for comparison with the
data.  This year, we have a program to observe a few globular clusters
over a range of ages and metallicities, and this program holds the
prospect of being able to make a comparison with the galaxies, leading to
an independent age for these systems.

\acknowledgments

We would like to thank Scott Trager, Guy Worthey, Bill Mathews, 
Rebecca Bernstein, Doug Richstone, Jimmy Irwin, and Tom Roellig 
for their valuable comments and advice. 
Particular thanks are due to Lorenzo Piovan, who provided his models
in digital form and was patient in explaining various details.
Also, the on-line model of Worthey, at
http://astro.wsu.edu/worthey/dial/dial\_a\_model.html were of great assistance.
We would like to thank NASA for their financial support of Spitzer 
program 3535.

\clearpage
\begin{deluxetable}{rrrrrrrrrrrrrrr}
\tablecolumns{15}
\tablewidth{0pc}
\tablecaption{Sample Galaxies and Optical Properties}
\tablehead{
\colhead{No.} & \colhead{Name} & \colhead{B$_0$} & \colhead{r$_e$} & \colhead{${\mu}_e$} & \colhead{$\sigma$} & \colhead{Z$_H$} & \colhead{Err} & \colhead{cz} & \colhead{M-m} & \colhead{M$_B$} & \colhead{logM} &  \colhead{M/L$_B$}  &  \colhead{Age}  &  \colhead{Err} \\
\colhead{} & \colhead{} & \colhead{mag} &\colhead{arcsec} & \colhead{} & \colhead{km s${-1}$} & \colhead{} & \colhead{} &\colhead{km s${-1}$} & \colhead{mag} & \colhead{mag} & \colhead{M$_\odot$} & \colhead{M$_\odot$/L$_\odot$} & \colhead{Gyr} & \colhead{Gyr}
}
\startdata
1  &  NGC 584  &  11.21  &  30  &  20.58  &  193  &  0.49  &  0.03  &  1866  &  31.6  &  -20.39  &  10.72  &  4.84  &  2.5  &  0.3 \\
2  &  NGC 636  &  12.22  &  19  &  20.72  &  160  &  0.34  &  0.07  &  1860  &  32.45  &  -20.23  &  10.53  &  4.04  &  4.1  &  0.7 \\
3  &  NGC 720  &  11.13  &  40  &  21.16  &  239  &  0.46  &  0.17  &  1741  &  32.29  &  -21.16  &  11.17  &  6.91  &  4.5  &  2.3 \\
4  &  NGC 821  &  11.72  &  36  &  21.49  &  189  &  0.23  &  0.03  &  1730  &  31.99  &  -20.27  &  10.86  &  7.47  &  7.5  &  1.2 \\
5  &  NGC 1339  &  12.5  &  17  &  20.64  &  158  &  0.12  &  0.07  &  1355  &  31.52  &  -19.02  &  10.28  &  6.28  &  12.7  &  4.8 \\
6  &  NGC 1351  &  12.48  &  26  &  21.33  &  157  &  -0.1  &  0.05  &  1529  &  31.52  &  -19.04  &  10.46  &  7.65  &  17  &  3.3 \\
7  &  NGC 1374  &  12.01  &  30  &  21.26  &  185  &  0.13  &  0.07  &  1349  &  31.52  &  -19.51  &  10.67  &  8.63  &  9.5  &  2.6 \\
8  &  NGC 1379  &  11.87  &  42  &  21.79  &  130  &  -0.08  &  0.06  &  1360  &  31.52  &  -19.65  &  10.51  &  4.96  &  10.9  &  2.9 \\
9  &  NGC 1399  &  10.44  &  42  &  20.68  &  375  &  0.29  &  0.06  &  1431  &  31.52  &  -21.08  &  11.43  &  14.85  &  11.5  &  2.4 \\
10  &  NGC 1404  &  10.98  &  27  &  20.02  &  260  &  0.25  &  0.05  &  1923  &  31.52  &  -20.54  &  10.92  &  6.04  &  9  &  2.5 \\
11  &  NGC 1427  &  11.81  &  33  &  21.34  &  175  &  -0.07  &  0.03  &  1416  &  31.52  &  -19.71  &  10.66  &  7.56  &  12.2  &  1.6 \\
12  &  NGC 1453  &  12.26  &  28  &  21.47  &  286  &  0.32  &  0.06  &  3886  &  33.59  &  -21.33  &  11.43  &  10.34  &  7.6  &  1.9 \\
13  &  IC 2006  &  12.25  &  29  &  21.45  &  136  &  0.06  &  0.06  &  1371  &  31.52  &  -19.27  &  10.39  &  5.75  &  16.9  &  4.2 \\
14  &  NGC 1700  &  12.01  &  24  &  20.82  &  227  &  0.5  &  0.03  &  3895  &  33.31  &  -21.3  &  11.11  &  4.75  &  2.3  &  0.3 \\
15  &  NGC 2300  &  11.77  &  34  &  21.42  &  252  &  0.38  &  0.05  &  1938  &  32.15  &  -20.38  &  11.12  &  12.25  &  5.9  &  1.5 \\
16  &  NGC 3377  &  11.07  &  34  &  20.78  &  108  &  0.2  &  0.06  &  724  &  30.33  &  -19.26  &  10.02  &  2.89  &  3.7  &  0.8 \\
17  &  NGC 3379  &  10.18  &  35  &  20.15  &  203  &  0.22  &  0.03  &  945  &  30.2  &  -20.02  &  10.55  &  5.88  &  8.6  &  1.4 \\
18  &  NGC 4472  &  9.33  &  104  &  21.4  &  279  &  0.26  &  0.05  &  980  &  31.14  &  -21.81  &  11.49  &  7.67  &  7.9  &  1.7 \\
19  &  NGC 4478  &  12.21  &  14  &  19.87  &  128  &  0.3  &  0.1  &  1365  &  31.37  &  -19.16  &  9.99  &  2.64  &  4.6  &  2.3 \\
20  &  NGC 4552  &  10.57  &  30  &  20.22  &  252  &  0.28  &  0.04  &  364  &  31.01  &  -20.44  &  10.83  &  7.77  &  10.5  &  1.2 \\
21  &  NGC 4649  &  9.7  &  74  &  21.11  &  310  &  0.29  &  0.04  &  1117  &  31.21  &  -21.51  &  11.45  &  9.87  &  11.7  &  1.5 \\
22  &  NGC 5638  &  12.06  &  34  &  21.58  &  154  &  0.2  &  0.03  &  1649  &  32.18  &  -20.12  &  10.7  &  5.23  &  8.3  &  1.4 \\
23  &  NGC 5812  &  11.83  &  22  &  20.65  &  200  &  0.39  &  0.04  &  1929  &  32.23  &  -20.4  &  10.74  &  5.65  &  5.3  &  1.1 \\
24  &  NGC 5813  &  11.42  &  49  &  21.83  &  205  &  -0.03  &  0.03  &  1954  &  32.62  &  -21.2  &  11.19  &  6.61  &  18.3  &  2.3 \\
25  &  NGC 5831  &  12.31  &  27  &  21.44  &  160  &  0.54  &  0.03  &  1655  &  32.25  &  -19.94  &  10.64  &  6.05  &  2.6  &  0.3 \\
26  &  NGC 5846  &  10.91  &  83  &  22.26  &  224  &  0.15  &  0.05  &  1714  &  32.06  &  -21.15  &  11.38  &  8.96  &  13.5  &  3.3 \\
27  &  NGC 6703  &  11.97  &  24  &  20.88  &  183  &  0.32  &  0.06  &  2403  &  32.18  &  -20.21  &  10.69  &  5.49  &  4.3  &  0.7 \\
28  &  NGC 7562  &  12.37  &  25  &  21.28  &  248  &  0.21  &  0.04  &  3608  &  33.87  &  -21.5  &  11.31  &  6.42  &  7.6  &  1.6 \\
29  &  NGC 7619  &  11.93  &  32  &  21.52  &  300  &  0.21  &  0.03  &  3762  &  33.7  &  -21.77  &  11.55  &  9.91  &  14.4  &  2.2 \\
\enddata
\end{deluxetable}

\clearpage
\thispagestyle{empty}
\begin{deluxetable}{rrrrrrrrrrrrrrrrrr}
\tablecolumns{18}
\tablewidth{0pc}
\tablecaption{Mid-Infrared Fluxes and Slopes}
\tablehead{
\colhead{No.} & \colhead{Name} & \colhead{$\alpha$(6$\mu$m)} & \colhead{Err} & \colhead{${\lambda}_{1}$} & \colhead{F$_{\nu ,1}$} & \colhead{Err} & \colhead{${\lambda}_{2}$} & \colhead{F$_{\nu ,2}$} & \colhead{Err} & \colhead{${\lambda}_{3}$}  & \colhead{F$_{\nu ,3}$} & \colhead{Err} & \colhead{${\lambda}_{4}$}  & \colhead{F$_{\nu ,4}$} & \colhead{Err} & \colhead{F$_{\nu ,3}$/F$_{\nu ,1}$} & \colhead{Err}
}
\startdata
1 & NGC 584 & -0.98 & 0.11 & 5.81 & 26.7 & 0.1 & 8.32 & 13.8 & 0.5 & 9.92 & 15.2 & 0.2 &
12.98 & 12.3 & 0.2 & 0.57 & 0.007 \\
2 & NGC 636 & -1.51 & 0.37 & 5.85 & 13.0 & 0.2 & 8.32 & 7.5 & 0.3 & 9.92 & 7.2 & 0.2 &
12.98 & 5.4 & 0.1 & 0.56 & 0.016 \\
3 & NGC 720 & -0.83 & 0.24 & 5.91 & 25.7 & 0.1 & 8.32 & 14.5 & 0.2 & 9.92 & 16.2 & 0.1 &
12.98 & 12.7 & 0.1 & 0.63 & 0.006 \\
4 & NGC 821 & -0.63 & 0.16 & 5.85 & 18.0 & 0.1 & 8.32 & 9.1 & 0.3 & 9.92 & 10.6 & 0.1 &
12.98 & 7.4 & 0.1 & 0.59 & 0.009 \\
5 & NGC 1339 & -0.73 & 0.31 & 5.82 & 15.5 & 0.2 & 8.32 & 8.6 & 0.2 & 9.92 & 8.9 & 0.1 &
12.98 & 6.3 & 0.1 & 0.57 & 0.011 \\
6 & NGC 1351 & -0.78 & 0.20 & 5.82 & 11.9 & 0.2 & 8.30 & 6.8 & 0.3 & 9.92 & 6.5 & 0.1 &
12.95 & 4.7 & 0.1 & 0.55 & 0.011 \\
7 & NGC 1374 & -0.79 & 0.22 & 5.82 & 15.6 & 0.1 & 8.32 & 9.3 & 0.2 & 9.92 & 9.2 & 0.1 &
12.98 & 7.1 & 0.1 & 0.59 & 0.007 \\
8 & NGC 1379 & -0.62 & 0.25 & 5.82 & 10.3 & 0.2 & 8.32 & 6.2 & 0.2 & 9.92 & 6.0 & 0.1 &
12.98 & 4.3 & 0.1 & 0.59 & 0.016 \\
9 & NGC 1399 & -0.91 & 0.10 & 5.82 & 42.5 & 0.2 & 8.32 & 21.7 & 0.2 & 9.92 & 25.3 & 0.1
& 12.98 & 19.8 & 0.1 & 0.59 & 0.004 \\
10 & NGC 1404 & -1.01 & 0.10 & 5.82 & 43.6 & 0.2 & 8.30 & 24.3 & 0.2 & 9.92 & 27.1 & 0.1
& 12.98 & 21.0 & 0.1 & 0.62 & 0.004 \\
11 & NGC 1427 & -0.74 & 0.25 & 5.82 & 13.1 & 0.2 & 8.30 & 7.3 & 0.2 & 9.92 & 8.0 & 0.1 &
12.98 & 5.4 & 0.1 & 0.61 & 0.011 \\
12 & NGC 1453 & -1.58 & 0.31 & 5.88 & 17.6 & 0.1 & 8.32 & 9.5 & 0.1 & 9.92 & 10.1 & 0.1
& 12.98 & 8.9 & 0.1 & 0.57 & 0.007 \\
13 & IC 2006 & -1.31 & 0.22 & 5.79 & 11.3 & 0.2 & 8.32 & 6.1 & 0.2 & 9.92 & 6.3 & 0.1 &
12.98 & 4.8 & 0.1 & 0.55 & 0.012 \\
14 & NGC 1700 & -0.76 & 0.19 & 5.83 & 18.3 & 0.2 & 8.35 & 11.4 & 0.0 & 9.92 & 12.1 & 0.2
& 12.98 & 8.7 & 0.1 & 0.66 & 0.011 \\
15 & NGC 2300 & -0.74 & 0.19 & 5.82 & 19.8 & 0.2 & 8.32 & 10.7 & 0.2 & 9.92 & 11.8 & 0.1
& 12.98 & 9.0 & 0.1 & 0.60 & 0.009 \\
16 & NGC 3377 & -1.04 & 0.23 & 5.85 & 26.9 & 0.2 & 8.32 & 15.8 & 0.3 & 9.92 & 18.0 & 0.1
& 12.98 & 12.8 & 0.1 & 0.67 & 0.007 \\
17 & NGC 3379 & -0.47 & 0.11 & 5.92 & 49.2 & 0.3 & 8.32 & 33.8 & 0.2 & 9.92 & 32.2 & 0.2
& 12.98 & 21.8 & 0.1 & 0.65 & 0.005 \\
18 & NGC 4472 & -1.04 & 0.07 & 5.82 & 48.0 & 0.2 & 8.32 & 25.7 & 0.2 & 9.92 & 29.4 & 0.1
& 12.98 & 22.8 & 0.1 & 0.61 & 0.003 \\
19 & NGC 4478 & -1.34 & 0.24 & 5.82 & 15.3 & 0.2 & 8.32 & 7.6 & 0.2 & 9.92 & 7.8 & 0.1 &
12.98 & 6.1 & 0.1 & 0.51 & 0.007 \\
20 & NGC 4552 & -1.19 & 0.14 & 5.82 & 45.6 & 0.2 & 8.32 & 24.5 & 0.3 & 9.92 & 26.9 & 0.2
& 12.98 & 21.4 & 0.1 & 0.59 & 0.004 \\
21 & NGC 4649 & -0.96 & 0.11 & 5.82 & 36.5 & 0.2 & 8.32 & 20.6 & 0.2 & 9.93 & 23.4 & 0.1
& 12.98 & 18.3 & 0.1 & 0.64 & 0.005 \\
22 & NGC 5638 & -0.66 & 0.28 & 5.82 & 12.1 & 0.2 & 8.32 & 7.2 & 0.2 & 9.92 & 7.4 & 0.2 &
12.97 & 5.2 & 0.1 & 0.61 & 0.016 \\
23 & NGC 5812 & -0.36 & 0.18 & 5.85 & 21.0 & 0.2 & 8.32 & 13.0 & 0.4 & 9.90 & 11.6 & 0.1
& 12.98 & 9.7 & 0.1 & 0.55 & 0.008 \\
24 & NGC 5813 & -0.79 & 0.11 & 5.82 & 21.0 & 0.1 & 8.29 & 10.4 & 0.1 & 9.93 & 11.3 & 0.1
& 12.98 & 8.5 & 0.1 & 0.54 & 0.007 \\
25 & NGC 5831 & -1.46 & 0.26 & 5.82 & 11.9 & 0.1 & 8.32 & 6.4 & 0.2 & 9.92 & 6.8 & 0.1 &
12.98 & 5.5 & 0.1 & 0.57 & 0.012 \\
26 & NGC 5846 & -1.03 & 0.18 & 5.82 & 18.9 & 0.2 & 8.32 & 12.0 & 0.2 & 9.92 & 12.1 & 0.1
& 12.98 & 9.5 & 0.1 & 0.64 & 0.007 \\
27 & NGC 6703 & -1.25 & 0.24 & 5.85 & 16.4 & 0.1 & 8.32 & 8.4 & 0.3 & 9.92 & 9.9 & 0.1 &
12.98 & 7.9 & 0.1 & 0.61 & 0.009 \\
28 & NGC 7562 & -0.49 & 0.31 & 5.82 & 13.9 & 0.1 & 8.32 & 7.9 & 0.2 & 9.92 & 7.5 & 0.1 &
12.98 & 4.8 & 0.1 & 0.54 & 0.008 \\
29 & NGC 7619 & -1.50 & 0.20 & 5.82 & 18.4 & 0.2 & 8.32 & 10.6 & 0.2 & 9.92 & 10.1 & 0.1
& 12.98 & 8.1 & 0.0 & 0.55 & 0.009 \\
\enddata
\end{deluxetable}

\clearpage

\begin{deluxetable}{rrrrrrrrr}
\tablecolumns{9}
\tablewidth{0pc}
\tablecaption{Derived Ages of Galaxies}
\tablehead{
\colhead{No.} & \colhead{Name} & \colhead{t(Trager)} & \colhead{Err} & \colhead{t(Thomas)} & \colhead{Err} & \colhead{M/L$_{B, cor}$} & \colhead{t(M/L$_B$)} & \colhead{t(IR)} \\
\colhead{} & \colhead{} & \colhead{Gyr} & \colhead{Gyr} & \colhead{Gyr} & \colhead{Gyr} & \colhead{M$_\odot$/L$_\odot$} & \colhead{Gyr} &\colhead{Gyr} \\
}
\startdata
1  &   584  &  2.5  &  0.3  &  2.8  &  0.3  &  5.6  &  7.0  &  14.6  \\
2  &   636  &  4.1  &  0.7  &  4.4  &  0.6  &  4.8  &  6.0  &  15.9  \\ 
3  &   720  &  4.5  &  2.3  &  5.4  &  2.4  &  6.7  &  8.4  &  10.8  \\ 
4  &   821  &  7.5  &  1.2  &  8.9  &  1.2  &  8.8  &  11.4  &  13.2  \\ 
5  &   1339  &  12.7  &  4.8  &     &     &  9.9  &  12.8  &  14.4  \\  
6  &   1351  &  17  &  3.3  &     &     &  12.0  &  15.7  &  16.4  \\
7  &   1374  &  9.5  &  2.6  &     &     &  12.2  &  15.9  &  13.4  \\
8  &   1379  &  10.9  &  2.9  &     &     &  6.8  &  8.6  &  13.4  \\
9  &   1399  &  11.5  &  2.4  &     &     &  14.6  &  19.3  &  12.9  \\  
10  &   1404  &  9  &  2.5  &     &     &  6.7  &  8.5  &  11.2  \\
11  &   1427  &  12.2  &  1.6  &     &     &  10.2  &  13.2  &  12.0  \\
12  &   1453  &  7.6  &  1.9  &  9.4  &  1.6  &  9.6  &  12.4  &  14.4  \\  
13  &  IC 2006  &  16.9  &  4.2  &     &     &  8.6  &  11.0  &  16.1  \\
14  &   1700  &  2.3  &  0.3  &  2.6  &  0.3  &  4.4  &  5.5  &  9.3  \\  
15  &   2300  &  5.9  &  1.5  &  7.3  &  1.5  &  14.1  &  18.7  &  12.8 \\  
16  &   3377  &  3.7  &  0.8  &  3.6  &  0.5  &  4.3  &  5.3  &  8.9  \\
17  &   3379  &  8.6  &  1.4  &  10  &  1.1  &  7.4  &  9.4  &  9.6  \\
18  &   4472  &  7.9  &  1.7  &  9.6  &  1.4  &  6.4  &  8.0  &  11.7  \\  
19  &   4478  &  4.6  &  2.3  &  5.3  &  1.2  &  4.0  &  5.0  &  20.6  \\ 
20  &   4552  &  10.5  &  1.2  &  12.4  &  1.5  &  8.8  &  11.4  &  13.2 \\  
21  &   4649  &  11.7  &  1.5  &  14.1  &  1.5  &  8.8  &  11.3  &  10.2  \\  
22  &   5638  &  8.3  &  1.4  &  9.5  &  0.9  &  6.4  &  8.1  &  12.0  \\
23  &   5812  &  5.3  &  1.1  &  6.5  &  1  &  6.5  &  8.2  &  16.0  \\
24  &   5813  &  18.3  &  2.3  &  16.6  &  2.2  &  6.3  &  8.0  &  17.4  \\
25  &   5831  &  2.6  &  0.3  &  3  &  0.4  &  7.7  &  9.9  &  14.4  \\
26  &   5846  &  13.5  &  3.3  &  14.2  &  2.2  &  8.7  &  11.1  &  10.1  \\
27  &   6703  &  4.3  &  0.7  &  4.8  &  0.8  &  6.6  &  8.3  &  12.2  \\
28  &   7562  &  7.6  &  1.6  &  8.6  &  1.3  &  5.7  &  7.2  &  17.6  \\ 
29  &   7619  &  14.4  &  2.2  &  15.4  &  1.4  &  8.3  &  10.6  &  16.4 \\ 
\enddata
\end{deluxetable}

\clearpage

\begin{figure}
\plotone{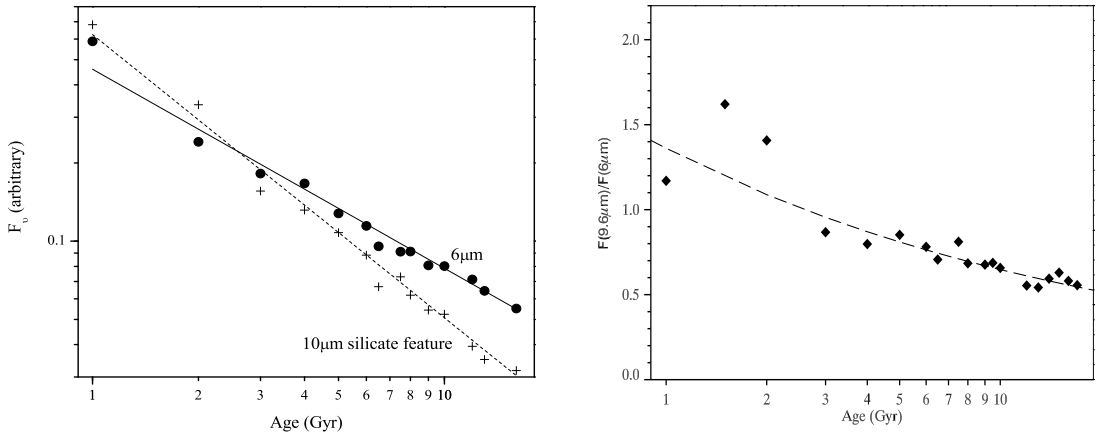}
\caption{Left Panel shows the decline of the 6 $\mu$m and the 9.6 $\mu$m 
luminosities
as a function of time for single-age populations, based on the solar 
metallicity models of \citet{piov03}, shown as data points.  Because 
the silicate feature
decreases more rapidly with time, the ratio of these luminosities forms
an age indicator. The evolution of the ratio $F(9.6\mu m)/F(6\mu m)$
as a function of age for a single stellar population with
metallicity Z=0.02 is presented in the righ panel. The dashed line represent
the relation between the flux ratio and the age t 
($t = 2.6\,(F(9.6\mu m)/F(6\mu m))^{-3.1}$ Gyr).  }
\end{figure}

\begin{figure}
\plotone{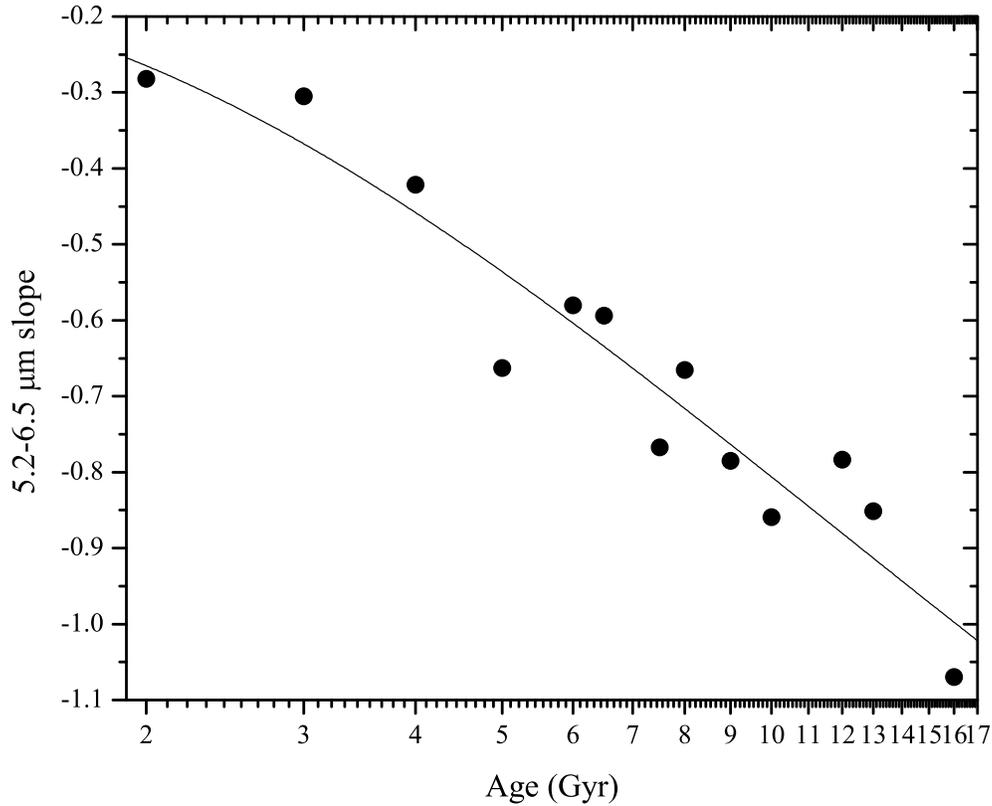}
\caption{The slope near 6 $\mu$m decreases with age in the solar metallicity
single-age populations of \citet{piov03}, shown as solid points.  A third-order 
polynomial is fit to the data.}
\end{figure}

\begin{figure}
\centering
\includegraphics{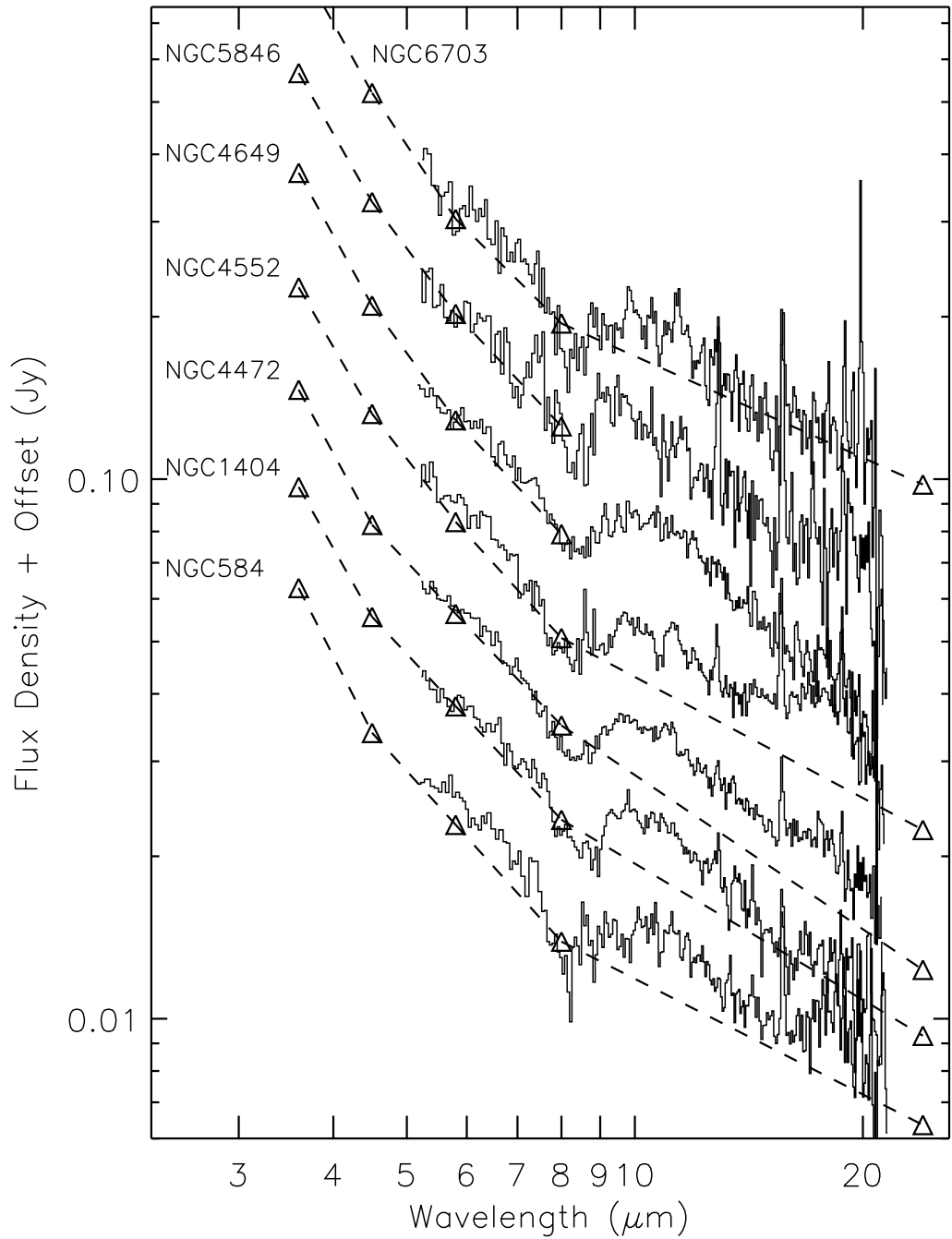}
\vskip.7in
\caption{
IRS low resolution spectra (solid line) of elliptical galaxies
are presented along with IRAC and MIPS broad-band photometric data
(triangles connected with dashed line). Spectra have been divided by
arbitrary factors for clarity and the broad-band data are scaled to match
the spectra at 8$\mu$m.  There is general agreement between the photometry
and the spectral energy densities.
}
\label{fig3}
\end{figure}

\begin{figure}
\plotone{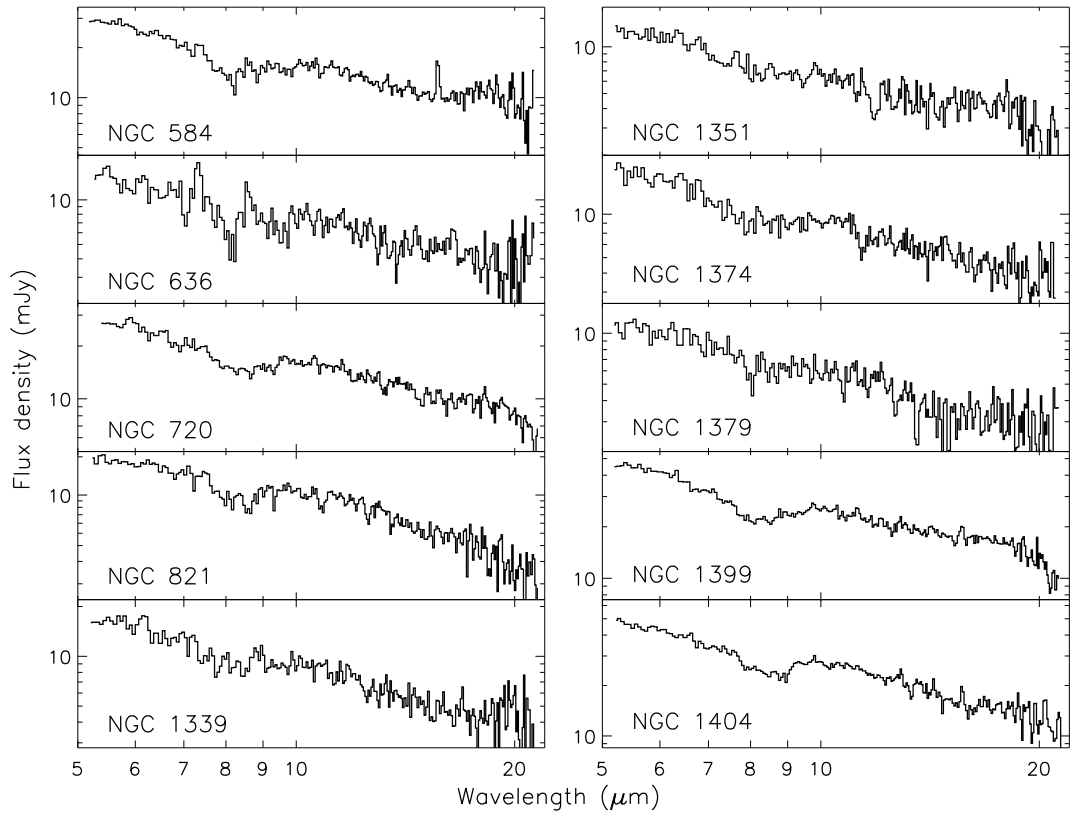}
\caption{The flux density as a function of wavelength
as obtained with the {\it IRS} on the {\it Spitzer Observatory}.  The
signal-to-noise is generally poor at the longest wavelengths.
}
\end{figure}

\clearpage
\plotone{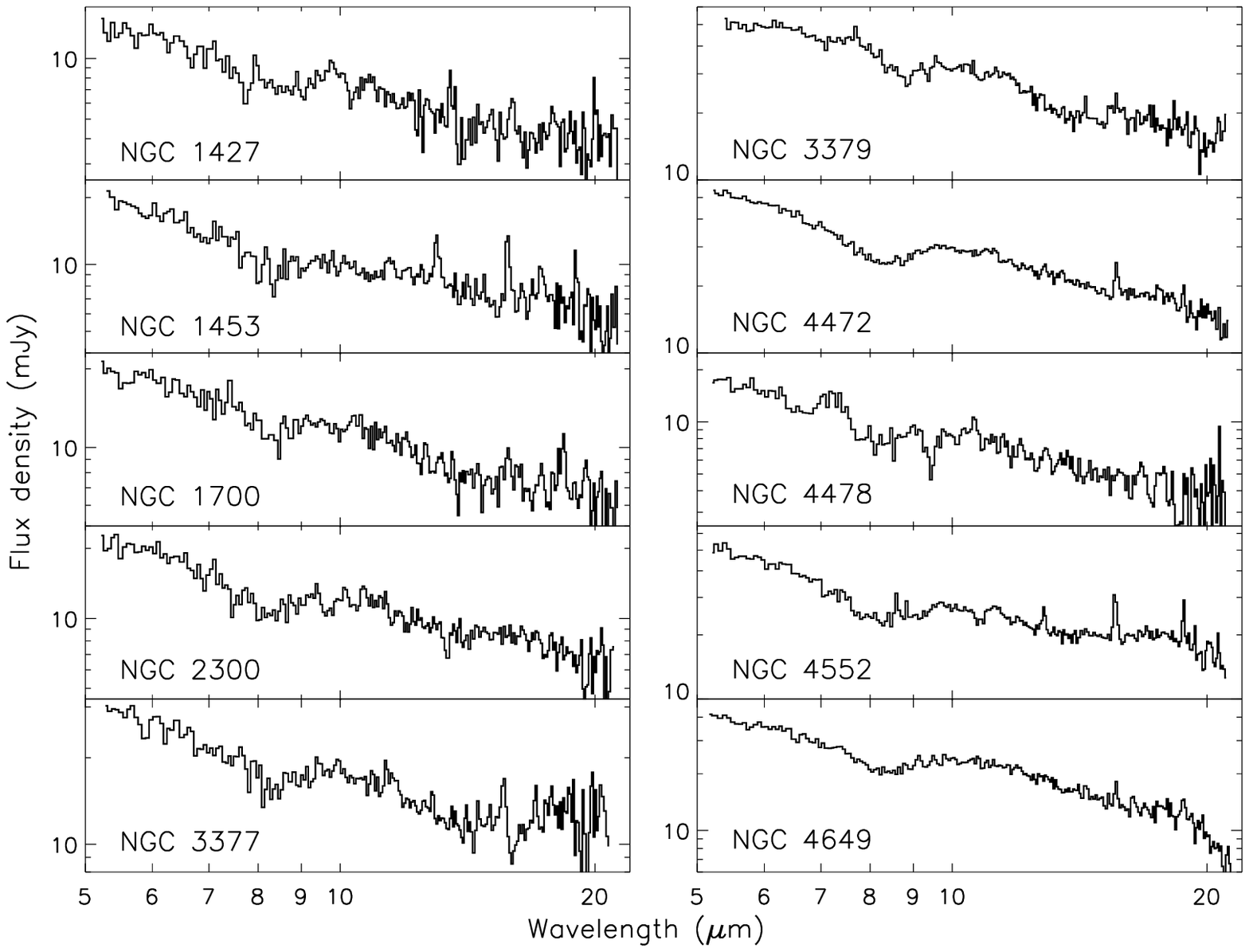}
\centerline{Fig. 4. --- Continued.}

\clearpage
\plotone{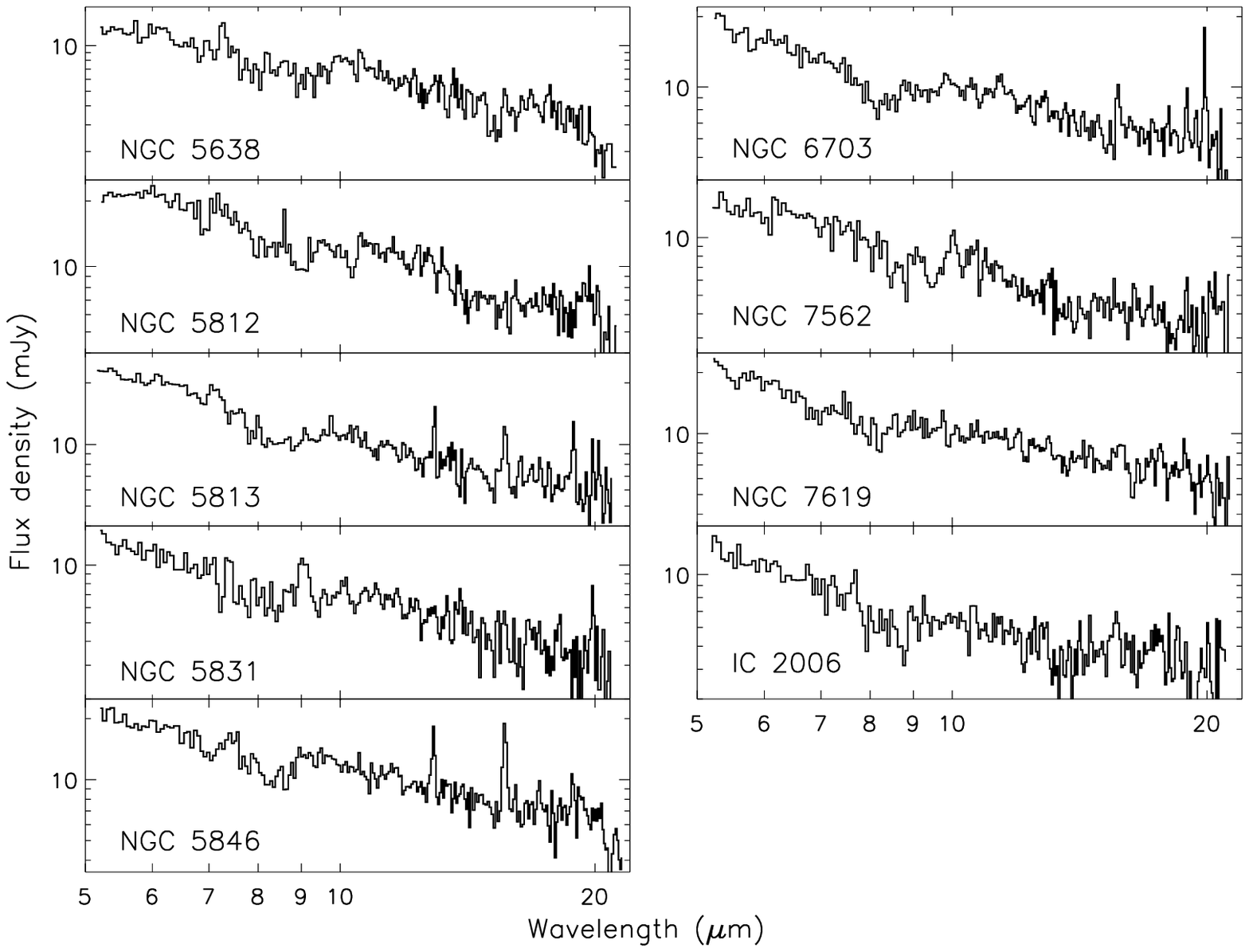}
\centerline{Fig. 4. --- Continued.}

\begin{figure}
\plottwo{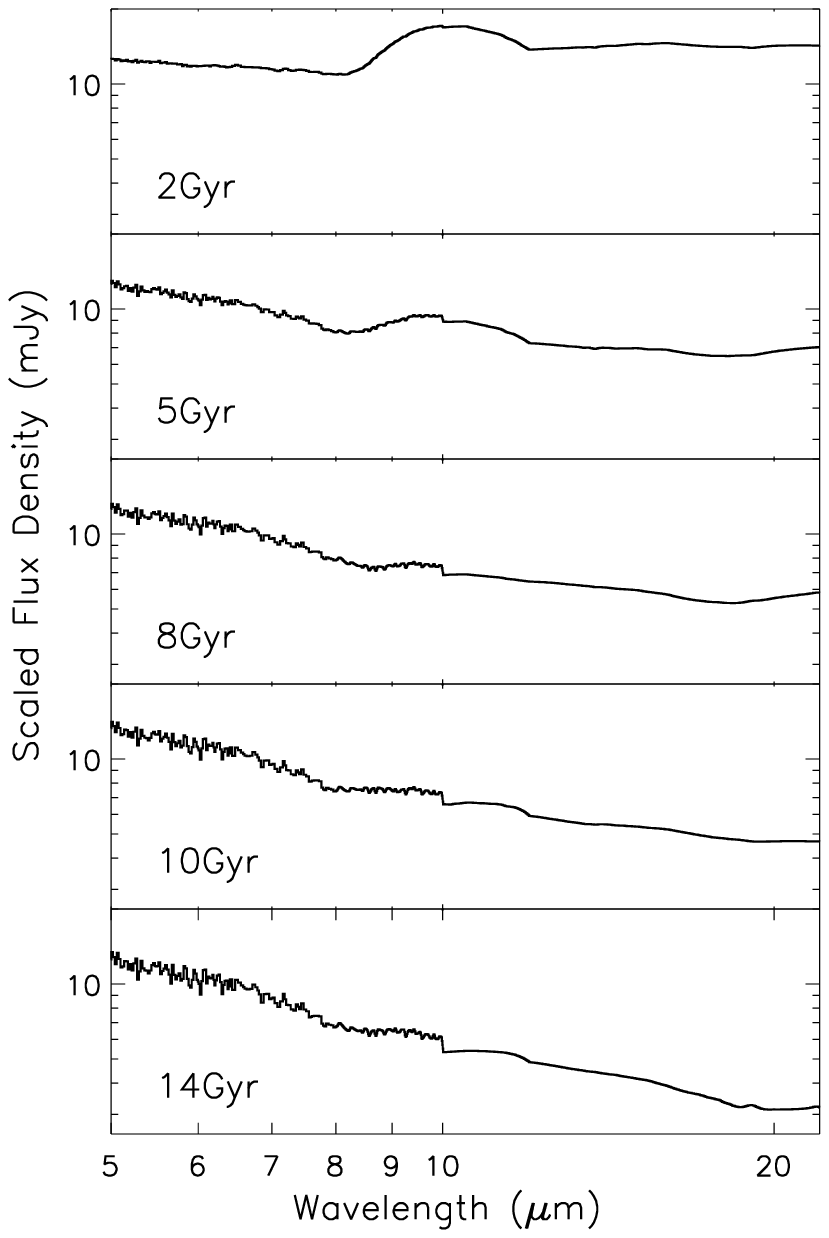}{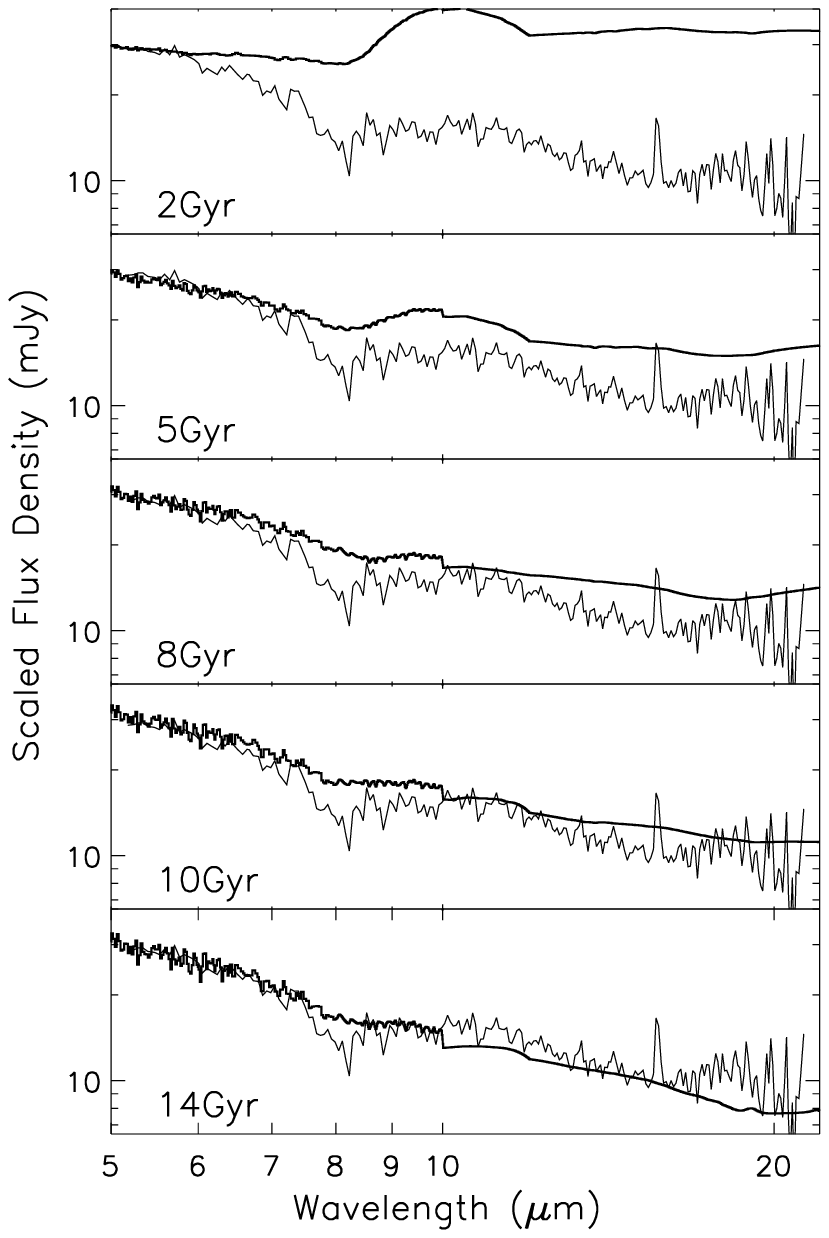}
\caption{The flux density of the spectral models of \citet{piov03} for
a single-age stellar population and at solar abundances (left panel) shows
the spectral steepening with age and the weakening of the 10 $\mu$m silicate
feature.  A comparison between these same models and the galaxy NGC 584
is shown in the right panel.  The best-fit appears to lie between 10 Gyr
and 14 Gyr, in conflict with the age derived from the optical absorption
line indices of 2.5 $\pm$ 0.3 Gyr.
}
\end{figure}

\begin{figure}
\plotone{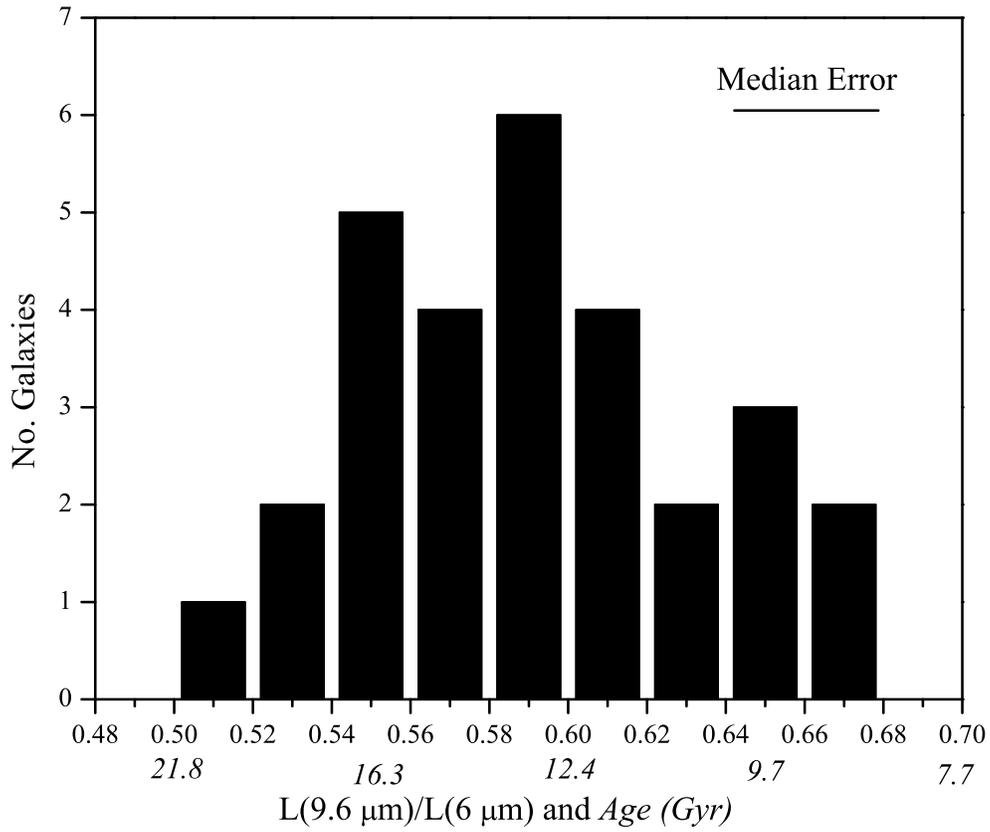}
\caption{The distribution of the ratio L(9.6 $\mu$m)/L(6 $\mu$m), with
the corresponding age, derived from \citet{piov03} along the bottom
in italics.  The mean ratio corresponds to an age of 13 Gyr and a 
$\sigma$ of about 3 Gyr, but much of that range is due to uncertainties
in the measurements, where the median error is shown.
}
\end{figure}

\begin{figure}
\plotone{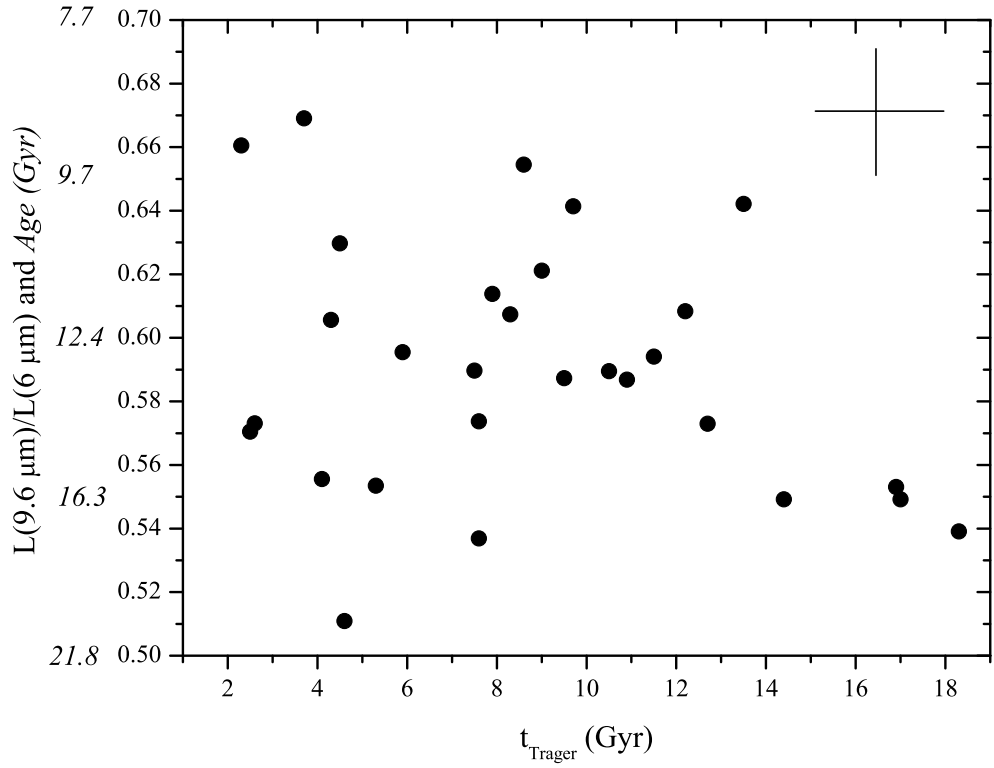}
\caption{The infrared indicator (and inferred ages in italic) vs the ages 
from \citet{trag00a} with average error bars in the upper right. The range
in ages from infrared data is much narrower than from the optical data and
there is no correlation between the two quantities.
}
\end{figure}

\begin{figure}
\plotone{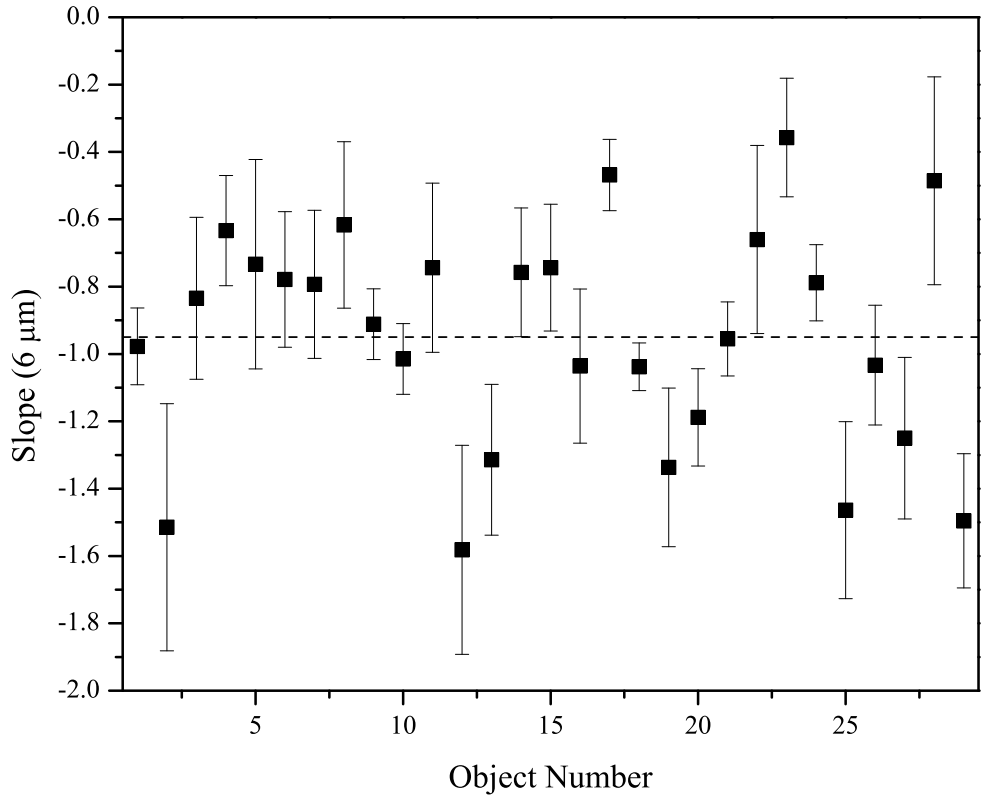}
\caption{The 6 $\mu$m spectral slope for objects by increasing RA, including
the statistical uncertainties and the mean value.  The mean value of -0.95 
(the dashed line) corresponds to 14 Gyr.  If systematic errors are smaller than
statistical errors, some variation in ages is required, a result driven by a
few objects with small errors.
}
\end{figure}

\begin{figure}
\plotone{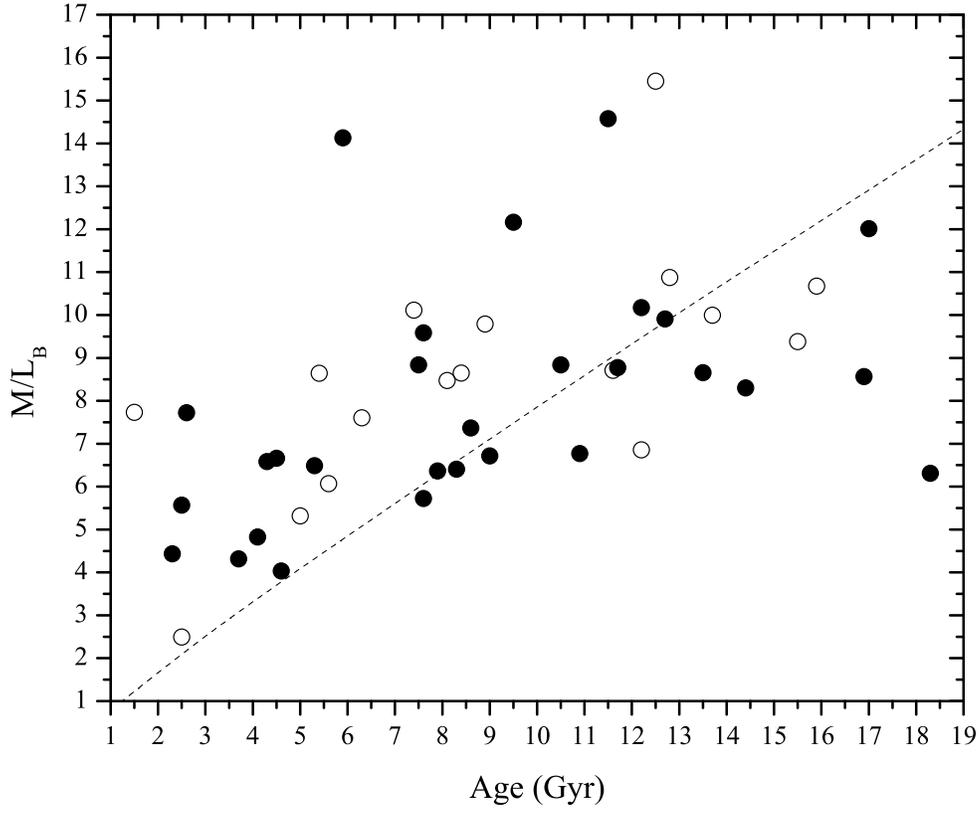}
\caption{The value of $M/L_B$ as a function of the ages from \citet{trag00a},
with the objects in our sample as black circles; $M/L_B$ has been corrected for
the luminosity dependence of the fundamental plane (normalized to M$_B$ = -21).  
The open circles are the additional objects given by \citet{trag00a}, 
with the exception of M31 and M32.  The solid line is the evolution 
of a single-age population with time.  While there are slightly more 
objects in the low-age, low-$M/L_B$ part of the figure, in general, there
is almost no relationship between $M/L_B$ and age.
}
\end{figure}

\begin{figure}
\plotone{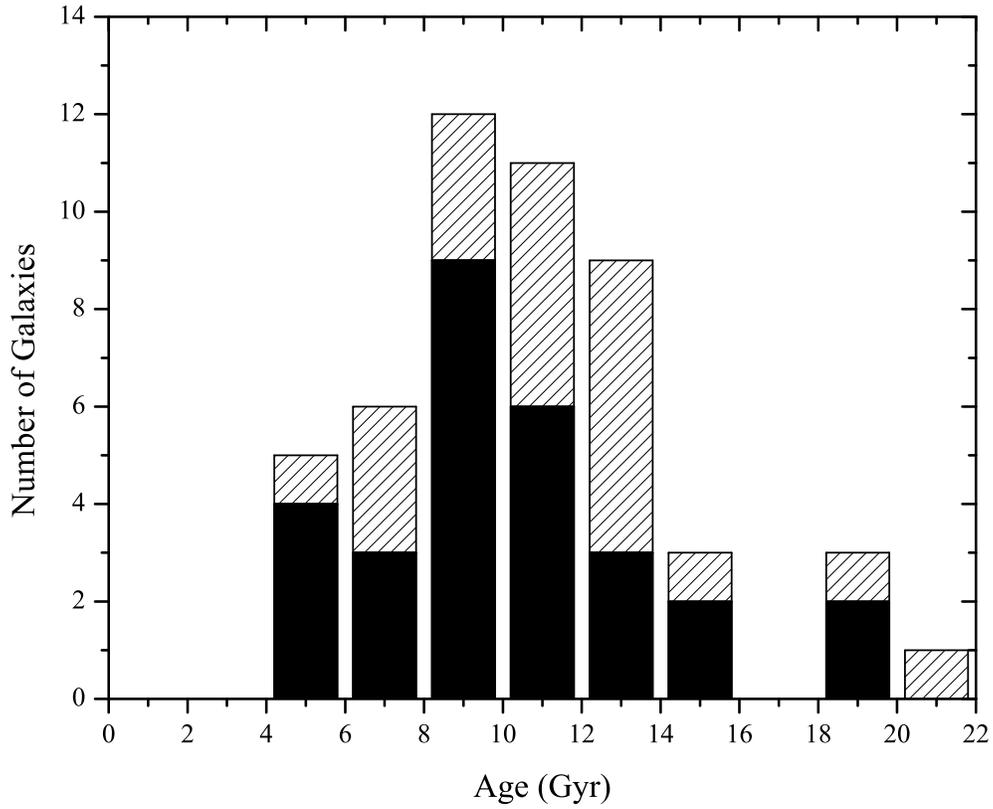}
\caption{The ages derived from $M/L_B$ given by \citet{trag00b}, 
after correcting it for the luminosity dependence of the fundamental plane.
The solid bars are the galaxies in this sample while the hatched regions
are the other objects in the list of \citet{trag00b}, with M31 and M32 excluded.
The distribution is similar to a Gaussian with a peak at 10.1 Gyr ($\pm$ 0.4 Gyr)
and $\sigma$ = 2.8 Gyr ($\pm$ 0.5 Gyr), excluding the outliers at 20 Gyr.
Most of this spread can be accounted for by measurement error.
}
\end{figure}

\begin{figure}
\plotone{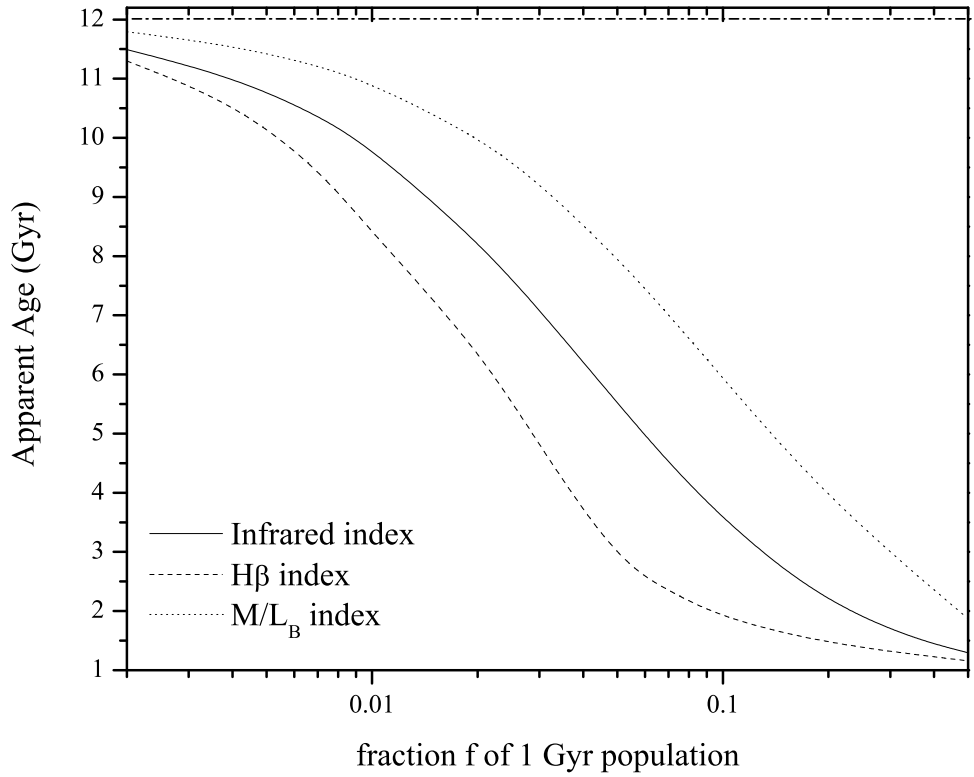}
\caption{Two populations are combined, the older one being 12 Gyr (dot-dash line
at top) and the other being a 1 Gyr population.  The Apparent Age is the value one
would infer by interpreting the mixed population as a single population.  Ages
inferred from the infrared index (solid line), the optical H$\beta $ line, and $M/L_B$
are shown for a solar metallicity model.
}
\end{figure}

\begin{figure}
\plotone{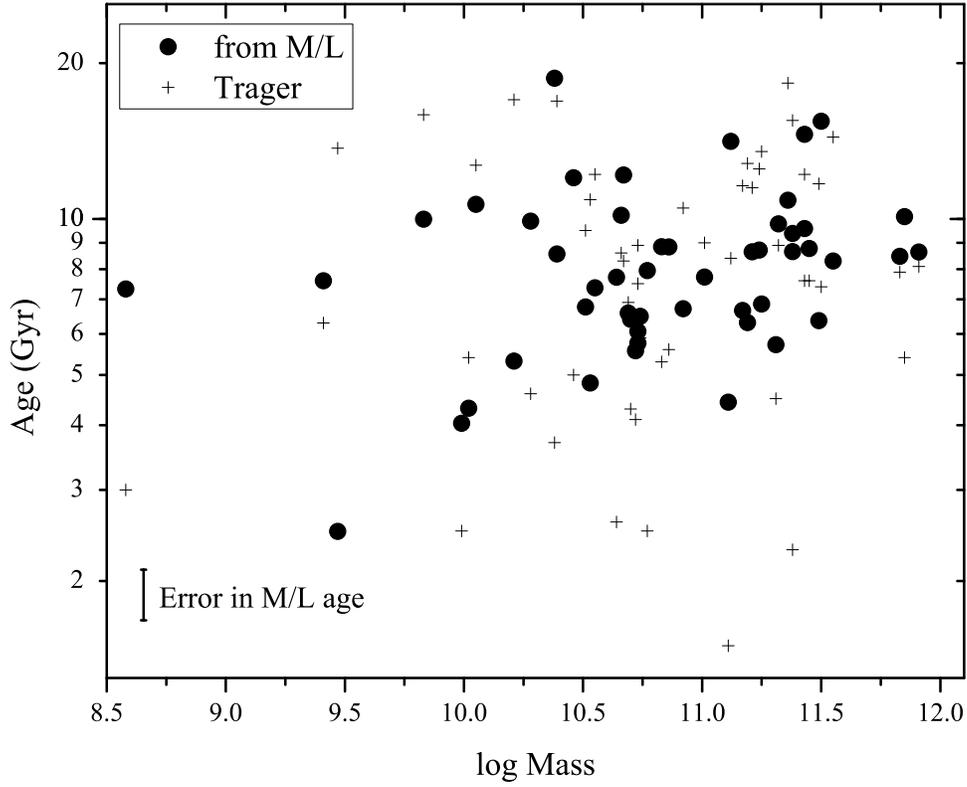}
\caption{The ages given by \citet{trag00a} and derived from $M/L_B$
as a function of galaxy mass for 49 galaxies in the original survey,
excluding M31 and M32.  Neither group shows a correlation with
mass.  Based on the Trager ages, young galaxies occur at all mass scales.
}
\end{figure}


\begin{thebibliography}{}
\bibitem[Bell et al.(2004)]{bell04} Bell, E.~F., et al.\ 2004,
\apj, 608, 752

\bibitem[Bell et al.(2005)]{bell05} Bell, E.~F., et al.\ 2005, 
\apj, 625, 23 

\bibitem[Bressan et al.(1998)]{bres98} Bressan, A., Granato, 
G.~L., \& Silva, L.\ 1998, \aap, 332, 135 

\bibitem[Brown et al.(2000)]{brow00} Brown, R.~J.~N., Forbes, 
D.~A., Kissler-Patig, M., \& Brodie, J.~P.\ 2000, \mnras, 317, 406 

\bibitem[Bruzual \& Charlot(2003)]{bruz03} Bruzual, G., \& 
Charlot, S.\ 2003, \mnras, 344, 1000 

\bibitem[Cappellari et al.(2005)]{capp05} Cappellari, M., et 
al.\ 2005, ArXiv Astrophysics e-prints, arXiv:astro-ph/0505042 

\bibitem[Cohen et al.(2003)]{cohen03} Cohen, M., Wheaton, 
W.~A., \& Megeath, S.~T.\ 2003, \aj, 126, 1090 

\bibitem[Daddi et al.(2005)]{dadd05a} Daddi, E., et al.\ 2005,
\apjl, 631, L13

\bibitem[Daddi et al.(2005)]{dadd05b} Daddi, E., et al.\ 2005,
\apj, 626, 680

\bibitem[Decin et al.(2004)]{decin04} Decin, L., Morris, P.~W., 
Appleton, P.~N., Charmandaris, V., Armus, L., \& Houck, J.~R.\ 2004, \apjs, 
154, 408 

\bibitem[Denicol{\'o} et al.(2005a)]{deni05a} Denicol{\'o}, G.,
Terlevich, R., Terlevich, E., Forbes, D.~A., Terlevich, A., \& Carrasco,
L.\ 2005, \mnras, 356, 1440

\bibitem[Denicol{\'o} et al.(2005b)]{deni05b} Denicol{\'o}, G.,
Terlevich, R., Terlevich, E., Forbes, D.~A., \& Terlevich, A.\ 2005,
\mnras, 358, 813

\bibitem[Faber et al.(2005)]{fab05} Faber, S.~M., et al.\
2005, ArXiv Astrophysics e-prints, arXiv:astro-ph/0506044

\bibitem[Higdon et al.(2004)]{higd04} Higdon, S.~J.~U., et 
al.\ 2004, \pasp, 116, 975 

\bibitem[Houck et al.(2004)]{houck04} Houck, J.~R., et al.\ 
2004, \apjs, 154, 18 

\bibitem[Kauffmann \& Charlot(1998)]{kauf98} Kauffmann, G., \& 
Charlot, S.\ 1998, \mnras, 294, 705 

\bibitem[Kronawitter et al.(2000)]{kron00} Kronawitter, A., 
Gerhard, O.~E., Saglia, R.~P., \& Bender, R.\ 2000, ASP Conf.~Ser.~197: 
Dynamics of Galaxies: from the Early Universe to the Present, 197, 99 

\bibitem[Labb{\'e} et al.(2005)]{labb05} Labb{\'e}, I., et 
al.\ 2005, \apjl, 624, L81 

\bibitem[Lan{\c c}on \& Mouhcine(2002)]{lanc02} Lan{\c c}on, 
A., \& Mouhcine, M.\ 2002, \aap, 393, 167 

\bibitem[Maraston \& Thomas(2000)]{mara00} Maraston, C., \&
Thomas, D.\ 2000, \apj, 541, 126

\bibitem[Meneux et al.(2005)]{mene05} Meneux, B., et al.\ 
2005, ArXiv Astrophysics e-prints, arXiv:astro-ph/0511656 

\bibitem[Mouhcine \& Lan{\c c}on(2003)]{mouh03} Mouhcine, M., 
\& Lan{\c c}on, A.\ 2003, \aap, 402, 425 

\bibitem[Piovan et al.(2003)]{piov03} Piovan, L., Tantalo, R., 
\& Chiosi, C.\ 2003, \aap, 408, 559 

\bibitem[Pipino \& Matteucci(2005)]{pipi05} Pipino, A., \& 
Matteucci, F.\ 2005, ArXiv Astrophysics e-prints,
arXiv:astro-ph/0510609 
 
\bibitem[Pipino \& Matteucci(2003)]{pipi03} Pipino, A., \& 
Matteucci, F.\ 2003, \apss, 284, 799 

\bibitem[Saglia et al.(2000)]{sagl00} Saglia, R.~P., 
Kronawitter, A., Gerhard, O., \& Bender, R.\ 2000, \aj, 119, 153 

\bibitem[Schweizer \& Seitzer(1992)]{schw92} Schweizer, F., \& 
Seitzer, P.\ 1992, \aj, 104, 1039 

\bibitem[Smith et al.(2004)]{smit04} Smith, J.~D.~T., et al.\ 
2004, \apjs, 154, 199 

\bibitem[Statler et al.(1996)]{stat96} Statler, T.~S., 
Smecker-Hane, T., \& Cecil, G.~N.\ 1996, \aj, 111, 1512 

\bibitem[Tantalo \& Chiosi (2004)]{tanta} 
Tantalo, R. \&  Chiosi, C.\ 2004, \mnras, 353, 917

\bibitem[Temi et al.(2005a)]{temi05a} Temi, P., Mathews, W.~G., 
\& Brighenti, F.\ 2005, \apj, 622, 235 

\bibitem[Temi et al.(2005b)]{temi05b} Temi, P., Brighenti, F., 
\& Mathews, W.~G.\ 2005, \apjl, 635 

\bibitem[Terlevich \& Forbes(2002)]{terl02} Terlevich, A.~I., 
\& Forbes, D.~A.\ 2002, \mnras, 330, 547 

\bibitem[Thomas et al.(2005a)]{thom05a} Thomas, D., Maraston, C.,
Bender, R., \& de Oliveira, C. M.\ 2005, \apj 621,673 

\bibitem[Thomas et al.(2005b)]{thom05b} Thomas, J., Saglia, 
R.~P., Bender, R., Thomas, D., Gebhardt, K., Magorrian, J., Corsini, E.~M., 
\& Wegner, G.\ 2005, \mnras, 360, 1355 

\bibitem[Trager et al.(2000)]{trag00a} Trager, S.~C., Faber, 
S.~M., Worthey, G., \& Gonz{\'a}lez, J.~J.\ 2000, \aj, 120, 165 
 
\bibitem[Trager et al.(2000)]{trag00b} Trager, S.~C., Faber, 
S.~M., Worthey, G., \& Gonz{\'a}lez, J.~J.\ 2000, \aj, 119, 1645 

\bibitem[Treu et al.(2005)]{treu05} Treu, T., et al.\ 2005, 
\apj, 633, 174

\bibitem[Werner et al.(2004)]{wern04} Werner, M.~W., et al.\ 
2004, \apjs, 154, 1 

\bibitem[Worthey(1994)]{worth94a} Worthey, G.\ 1994, \apjs, 95, 
107 
 
\bibitem[Worthey et al.(1994)]{worth94b} Worthey, G., Faber, 
S.~M., Gonzalez, J.~J., \& Burstein, D.\ 1994, \apjs, 94, 687 

\bibitem[Whitmore et al.(1997)]{whit97} Whitmore, B.~C., 
Miller, B.~W., Schweizer, F., \& Fall, S.~M.\ 1997, \aj, 114, 1797 

\bibitem[Yi et al.(2005)]{yi05} Yi, S.~K., et al.\ 2005,
\apjl, 619, L111

\end{thebibliography}
\end{document}